\documentclass[nofootinbib,preprint,tightenlines,superscriptaddress]{revtex4}

\usepackage[utf8]{inputenc}
\usepackage[english]{babel}
\usepackage{amssymb,amsthm,amsmath,amstext,amsbsy,amsopn}
\usepackage{bbm}
\usepackage{nicefrac}
\usepackage{slashed}
\usepackage{pstricks}
\usepackage{graphicx}
\usepackage{hyperref}
\usepackage{leftidx}
\usepackage{environ}
\usepackage{mathtools}
\usepackage{xspace}
\usepackage{overpic}

\newcommand{\ie}{\textit{i.e.}}
\newcommand{\eg}{\textit{e.g.}}
\newcommand{\cf}{\textit{cf.}\xspace}

\newcommand{\etal}{\textit{et al.}\xspace}

\newcommand{\mathspace}{\ \ }
\newcommand{\mathtext}[1]{\mathspace\text{#1}\mathspace}

\newcommand{\fm}{\ensuremath{\mathrm{fm}}}

\newcommand{\vecr}{\mathbf{r}}

\newcommand{\vecp}{\mathbf{p}}

\newcommand{\vecq}{\mathbf{q}}

\newcommand{\vZero}{\mathbf{0}}

\newcommand{\vdelta}{\delta^{(3)}}

\newcommand{\dd}{\mathrm{d}}

\newcommand{\dq}[1]{\!\!\frac{\mathrm{d}^3#1}{(2\pi)^3}}

\newcommand{\dqd}[1]{\!\!\frac{\mathrm{d}^d#1}{(2\pi)^d}}

\newcommand{\ddq}{\dq{q}}

\newcommand{\gsim}{\gtrsim}
\newcommand{\lsim}{\lesssim}

\newcommand{\ii}{\mathrm{i}}
\newcommand{\ee}{\mathrm{e}}
\newcommand{\vD}{\boldsymbol{D}}
\newcommand{\hc}{\mathrm{h.c.}}
\newcommand{\OO}{\mathcal{O}}

\newcommand{\eps}{\varepsilon}

\newcommand{\Laplace}{\Delta}

\newcommand{\EulerGamma}{C_E}

\newcommand{\Rp}{\mathrm{Re}}
\newcommand{\Ip}{\mathrm{Im}}

\newcommand{\Si}{\mathrm{Si}}
\newcommand{\Ci}{\mathrm{Ci}}

\newcommand{\LiTwo}{\mathrm{Li}_2}

\newcommand{\mbraket}[3]{\langle #1|#2|#3\rangle}

\newcommand{\abs}[1]{\left|#1\right|}

\newcommand{\EB}{-E_B}
\newcommand{\EBp}{E_B}

\newcommand{\MN}{M_N}

\newcommand{\Mpi}{M_\pi}

\newcommand{\yd}{y_d}
\newcommand{\yt}{y_t}

\newcommand{\sigmad}{\sigma_d}

\newcommand{\sigmatpp}{\sigma_{t,pp}}
\newcommand{\sigmatnn}{\sigma_{t,nn}}

\newcommand{\cd}{c_d}
\newcommand{\ct}{c_t}
\newcommand{\ctpp}{c_{t,pp}}
\newcommand{\ctnn}{c_{t,nn}}

\newcommand{\gamd}{\gamma_d}

\newcommand{\rd}{\rho_d}
\newcommand{\rnt}{r_t}

\newcommand{\app}{a_C}
\newcommand{\rpp}{r_C}

\newcommand{\andDoublet}{\ensuremath{{}^2a_{\text{$n$--$d$}}}}

\newcommand{\Triton}{\ensuremath{{}^3\mathrm{H}}\xspace}
\newcommand{\He}{\ensuremath{{}^3\mathrm{He}}\xspace}

\newcommand{\LamNoPi}{\Lambda_{\slashed\pi}}

\newcommand*\rvec[1]%
{\ensuremath{\overset{\smash{\raisebox{-1.5pt}{\tiny$\rightarrow$}}}{#1}}}
\newcommand*\lvec[1]%
{\ensuremath{\overset{\smash{\raisebox{-1.5pt}{\tiny$\leftarrow$}}}{#1}}}

\newcommand{\vk}{\mathbf{k}}
\newcommand{\vp}{\mathbf{p}}
\newcommand{\vq}{\mathbf{q}}

\newcommand{\Tgen}{\mathcal{T}}
\newcommand{\Bgen}{\mathcal{B}}

\newcommand{\sss}{\mathrm{s}}

\newcommand{\BS}{\Bgen_\sss}
\newcommand{\BSda}{\BS^\mathrm{d,a}}
\newcommand{\BSdb}{\BS^\mathrm{d,b1}}
\newcommand{\BSdc}{\BS^\mathrm{d,b2}}

\newcommand{\TS}{\Tgen_\sss}
\newcommand{\TSda}{\TS^\mathrm{d,a}}
\newcommand{\TSdb}{\TS^\mathrm{d,b}}

\newcommand{\KS}{K_\sss}

\newcommand{\keV}{\ensuremath{\mathrm{keV}}}
\newcommand{\MeV}{\ensuremath{\mathrm{MeV}}}

\newcommand{\Tr}{\mathrm{Tr}}

\newcommand{\NLO}{\text{NLO}\xspace}
\newcommand{\NNLO}{\text{N$^2$LO}\xspace}

\newcommand{\skrowspace}{0.5em}

\NewEnviron{subalign}[1][]{%
\begin{subequations}\begin{align}
  \BODY
\end{align}\label{#1}\end{subequations}
}

\NewEnviron{spliteq}{%
\begin{equation}\begin{split}
  \BODY
\end{split}\end{equation}
}

\newcommand*{\vcenteredhbox}[1]
{\begingroup\setbox0=\hbox{#1}\parbox{\wd0}{\box0}\endgroup}

\begin{document}

\title{Effective Theory of $^3$H and $^3$He}

\author{Sebastian König}
\email{koenig.389@osu.edu}
\affiliation{Department of Physics, The Ohio State University,
Columbus, Ohio 43210, USA}

\author{Harald W. Grießhammer}
\email{hgrie@gwu.edu}
\affiliation{Institute for Nuclear Studies, Department of 
Physics, George Washington University, Washington DC 20052, USA}

\author{H.-W. Hammer}
\email{Hans-Werner.Hammer@physik.tu-darmstadt.de}
\affiliation{Institut für Kernphysik, Technische Universität Darmstadt, 
64289 Darmstadt, Germany}
\affiliation{ExtreMe Matter Institute EMMI, GSI Helmholtzzentrum 
für Schwerionenforschung GmbH, 64291 Darmstadt, Germany}

\author{U. van Kolck}
\email{vankolck@ipno.in2p3.fr}
\affiliation{Institut de Physique Nucléaire, CNRS/IN2P3, 
Université Paris-Sud, 91406 Orsay, France}
\affiliation{Department of Physics, University of Arizona,
Tucson, AZ 85721, USA}

\date{\today}

\begin{abstract}
We present a new perturbative expansion for pionless effective field theory with 
Coulomb interactions in which at leading order the spin-singlet 
nucleon--nucleon channels are taken in the unitarity limit.  Presenting results 
up to next-to-leading order for the Phillips line and the neutron--deuteron 
doublet-channel phase shift, we find that a perturbative expansion in the 
inverse ${}^1S_0$ scattering lengths converges rapidly.  Using a new 
systematic treatment of the proton--proton sector that isolates the divergence 
due to one-photon exchange, we renormalize the corresponding contribution to the 
\Triton--\He binding energy splitting and demonstrate that the Coulomb force in 
pionless EFT is a completely perturbative effect in the trinucleon bound-state 
regime.  In our new expansion, the leading order is exactly isospin-symmetric.  
At next-to-leading order, we include isospin breaking via the Coulomb force and 
two-body scattering lengths, and find for the energy splitting 
$(\EBp(\He)-\EBp(\Triton))^{\text{NLO}} = (-0.86 \pm 0.17)~\MeV$.
\end{abstract}

\maketitle

\section{Introduction}

It has been known for a long time that two-nucleon ($N\!N$) scattering
at very low energies can be described by the effective range 
expansion~\cite{Bethe:1949yr}, from which deuteron and $^1S_0$ virtual state 
properties emerge without information about details of the strong 
interaction~\cite{BethePeierls:1935}.  At these energies effects from 
pion-exchange physics cannot be resolved.  It is therefore possible to describe 
such a system using only nonrelativistic nucleons as degrees of freedom that 
interact via short-range (contact) 
forces~\cite{Bedaque:1997qi,vanKolck:1997ut,Kaplan:1998tg,Bedaque:1998mb,
Kaplan:1998we,Birse:1998dk,vanKolck:1998bw,Chen:1999tn,Bedaque:1999vb,
Gabbiani:1999yv}.  The systematic approach to implement this procedure, known as 
\emph{pionless effective field theory} (pionless EFT), is based on the 
experimental fact that the $N\!N$ $S$-wave scattering lengths are much larger 
than the corresponding effective ranges, so that a nonperturbative resummation 
of non-derivative two-body contact interactions is required at leading order 
(LO) to reproduce the shallow $N\!N$ bound and virtual states.

The extension of these ideas to the triton ($^3$H) and helion ($^3$He) was not 
immediate because a system of three nucleons collapses under the sole effect of 
attractive, non-derivative contact forces~\cite{Thomas:1935zz}.  The solution 
within pionless EFT is the existence of a non-derivative three-body 
interaction~\cite{Bedaque:1999ve,Hammer:2000nf,Hammer:2001gh,Bedaque:2002yg,
Afnan:2003bs,Griesshammer:2005ga} at LO.  This force provides not only 
saturation, but also a three-body parameter which correlates certain observables 
such as the triton binding energy and the doublet neutron-deuteron ($nd$) 
scattering length (``Phillips line''~\cite{Phillips:1968zze}).  This framework 
recovers Efimov's universal approach to the three-nucleon 
problem~\cite{Efimov:1970zz,Efimov:1981aa,Hammer:2010kp}.  With recent progress, 
pionless EFT also allows for an elegant and efficient fully perturbative 
treatment of contributions beyond 
LO~\cite{Hammer:2001gh,Vanasse:2013sda,Konig:2013cia,Vanasse:2014kxa}.

An important question is how far up the nuclear chart this EFT applies.  Nuclear 
density tends to increase with nucleon number $A$, implying larger typical 
nucleon momenta within the nucleus.  Calculations suggest that pionless EFT 
holds for 
$A=4$~\cite{Platter:2004zs,Kirscher:2009aj,Kirscher:2011uc,Kirscher:2015ana} 
and perhaps up to $A=6$ systems~\cite{Kirscher:2015ana,Stetcu:2006ey} without a 
four-body force at LO.  As the pion mass increases, its range of applicability 
increases, and this framework has been established as a powerful tool that can 
be used to analyze and supplement calculations of light nuclei directly from 
lattice QCD~\cite{Barnea:2013uqa,Beane:2015yha,Kirscher:2015yda}.  (See 
Refs.~\cite{Braaten:2003eu,Epelbaum:2006jc,Hammer:2007kq} for earlier work on 
pionless EFT for unphysical quark masses using input from chiral potentials.)

The application of EFT to nuclei requires an understanding of the importance of 
Coulomb and other electromagnetic effects.  The long-range nature of the 
Coulomb force implies that it becomes dominant at very low energies, \ie, 
precisely where the EFT is supposed to work best.  To describe scattering in 
this regime, one thus has to resum Coulomb effects to all orders to recover the 
Coulomb-modified effective-range expansion.  In the proton-proton ($pp$) sector 
of the pionless theory this was first carried out by Kong and
Ravndal~\cite{Kong:1998sx,Kong:1999sf}, with subsequent discussions of the 
charged two-body sector given in 
Refs.~\cite{Kong:2000px,Butler:2001jj,Barford:2002je,Ando:2007fh,Ando:2008va,
Ando:2008jb}.  An early attempt at $pd$ scattering was made in 
Ref.~\cite{Rupak:2001ci}, and extended to lower center-of-mass energies in 
Ref.~\cite{Konig:2011yq}.  A calculation of the Coulomb-modified $pd$ 
scattering length, with the consistent use of a Yukawa-screened Coulomb 
potential in momentum space, has been presented in Ref.~\cite{Konig:2013cia}.

The \Triton--\He binding energy difference has been studied using pionless EFT 
in a number of papers. Systems of three and four nucleons including the 
Coulomb interaction as part of an effective potential were first analyzed by 
Kirscher~\etal using the resonating group 
method~\cite{Kirscher:2009aj,Kirscher:2011uc,Kirscher:2011zn,Kirscher:2015ana}.
Ando and Birse~\cite{Ando:2010wq} carried out a momentum-space LO calculation 
which was nonperturbative both in the sense that it looked for the bound state 
as a pole in the $pd$ doublet-channel amplitude, as well as in the 
electromagnetic sector, where Coulomb effects were included through a fully 
off-shell Coulomb T-matrix, using methods developed by Kok~\etal\ in 
Refs.~\cite{Kok:1979aa,Kok:1981aa}.  A subset of the present authors presented 
a calculation of $pd$ scattering and the bound-state regime in 
Refs.~\cite{Konig:2011yq,Koenig:2013,Hammer:2014rba}.  
Of these, Refs.~\cite{Konig:2011yq,Koenig:2013} included a perturbative 
framework using trinucleon wave functions.  An updated version of this 
calculation that corrects some issues of the previous approach has recently been 
given in Ref.~\cite{Konig:2014ufa}.  That work also includes a nonperturbative 
treatment, in which all $\OO(\alpha)$ Coulomb diagrams are resummed to all 
orders.  This calculation was similar to that of Ref.~\cite{Ando:2010wq} but 
found that the full Coulomb T-matrix is not necessary for an accurate 
description of the \He bound state.

At LO in the strong interactions, the perturbative and nonperturbative results 
of Ref.~\cite{Konig:2014ufa} were found to agree with each other (as well 
as with the experimental value for the \Triton--\He binding energy difference)  
to within roughly 30\%.  While at first sight this seems fine if one keeps in 
mind the expected uncertainty based on the EFT expansion, the relevant 
parameter in this case is actually not the $Q/\LamNoPi$ of the strong sector 
(with the typical low-momentum scale set by the deuteron binding momentum, 
$Q\sim\gamma_d \simeq 45~\MeV$ and the breakdown scale $\LamNoPi\sim\Mpi\simeq 
140~\MeV$), but rather the $\alpha\MN/Q$ scale (with $\MN\simeq940~\MeV$) 
set by the Coulomb contributions.  Since in the bound-state regime the momentum 
scale is set by the trinucleon binding momentum, $\gamma_T \sim 80~\MeV$, 
Coulomb effects are expected to be a small perturbation.  Based on this, one 
should expect better agreement between perturbative and nonperturbative 
calculations.

The purpose of this work is to investigate a rearrangement of pionless EFT that 
explores the role of the increasing nuclear binding momentum in the simplest 
context, the trinucleon systems.  We develop a new formulation to include 
Coulomb and other electromagnetic effects in nuclear ground states using 
perturbation theory.  There are two important ingredients to this procedure.  
First, we introduce a new counting scheme that takes as LO the spin-singlet 
channel in the unitarity limit (infinite scattering length) and only includes 
the finite ${}^1S_0$ scattering length as a perturbative correction.  We will 
show with the \Triton binding energy and with doublet $nd$ scattering phase 
shifts that deviations from ${}^1S_0$ unitarity are indeed small.  
In the case of $pp$ scattering at very low energies, the scattering-length term
is iterated along with Coulomb effects so that renormalization can be carried 
out by matching to the Coulomb-modified effective-range expansion.  
This is essentially what was introduced in Ref.~\cite{Kong:1999sf}, but here 
we isolate the contribution from the divergent single-photon bubble, which was 
missed in the perturbative calculation of Ref.~\cite{Konig:2014ufa}.  
This new scheme is the second ingredient which then allows us to use the 
counterterm fixed by $pp$ scattering also in the perturbative calculation of the 
\Triton--\He binding energy difference at next-to-leading order (NLO) in the new 
counting.  The result differs from that of the nonperturbative calculation by 
much less than the 30\% previously obtained, and is slightly closer to the 
experimental value of the binding energy difference.  At NLO we end up about 
1.5\% off the \He binding energy.

Besides finally obtaining a complete calculation of the energy splitting that 
includes Coulomb effects only as perturbative corrections, we also employ a 
fully perturbative treatment of corrections in the strong sector.  This was 
already done in Ref.~\cite{Vanasse:2014kxa}, which showed that a new,
isospin-breaking three-body counterterm is needed for renormalization in the 
presence of range corrections when two-body Coulomb effects are resummed to all 
orders.  A further motivation for the current paper is to revisit this issue 
with a completely perturbative inclusion of those contributions as 
well.\footnote{The viability of perturbative Coulomb exchange for the 
trinucleon system was also investigated in independent 
work~\cite{Kirscher:2015zoa}, which appeared shortly after submission of our 
manuscript.}

This paper is structured as follows.  In Sec.~\ref{sec:Formalism} we discuss 
the basic formalism of pionless EFT with explicit isospin breaking.  The 
two-body sector and in particular our new method to treat the $pp$ channel is 
described in detail in Sec.~\ref{sec:TwoBody}.  In Sec.~\ref{sec:ThreeBody} we 
discuss the implications for the three-body sector.  The new results for the  
\Triton and \He binding energies are presented in Sec.~\ref{sec:Results}.  
We conclude in Sec.~\ref{sec:Conclusion} and provide technical details about 
the divergent Coulomb-bubble diagram in the Appendix.

\section{Effective Lagrangian}
\label{sec:Formalism}

We are interested in describing nuclear bound states in terms of the most 
general dynamics consistent with QCD symmetries built from a nucleon field $N$ 
of mass $M_N$.  We use a modified and extended version of the notation and 
conventions in Refs.~\cite{Koenig:2013,Konig:2014ufa}.  To NLO we write the 
Lagrangian as
\begin{equation}
 \mathcal{L} = N^\dagger\left(\ii D_0+\frac{\vD^2}{2\MN}\right)N
 +\mathcal{L}_\mathrm{2d}+\mathcal{L}_\mathrm{2t}+\mathcal{L}_3
\,,
\label{eq:L-Nd}
\end{equation}
where $D_\mu = \partial_\mu + \ii eA_\mu \hat{Q}_N$ includes the direct nucleon 
coupling to the electromagnetic field via the charge operator 
$\hat{Q}_N=(1+\tau_3)/2$.

It is convenient to express the interaction terms via auxiliary dibaryon
fields $d^i$ and $t^A$ in the $N\!N$ channels projected with
\begin{equation}
 P^i_d = \sigma^2\sigma^i\tau^2 / \sqrt8 \mathtext{,}
 P^A_t = \sigma^2\tau^2\tau^A/\sqrt8 \,,
\end{equation}
respectively, where lowercase (uppercase) letters are spin-$1$ (isospin-$1$) 
indices.  While $\hat{Q}_N$ ensures that Coulomb photons couple only to 
protons, the isospin-$1$ basis used for the spin-singlet field $t^A$ otherwise 
mixes contributions from protons and neutrons.  If we take $A=1,2,3$ to be an 
index in the Cartesian basis, then the $np$ configurations are completely 
contained in the $A=3$ component, whereas $nn$ and $pp$ are obtained from 
linear combinations $1\pm\ii2$.  This corresponds to using a spherical basis 
$\tilde{A} = -1,0,1$, where one would immediately have the desired separation.
We thus define new projectors
\begin{equation}
 \tilde{P}^{-1}_t = \frac{1}{\sqrt{2}}\left(P_t^1-\ii P_t^2\right)
 \mathtext{,}
 \tilde{P}^{0}_t = P_t^3
 \mathtext{,}
 \tilde{P}^{+1}_t = {-}\frac{1}{\sqrt{2}}\left(P_t^1+\ii P_t^2\right)
\end{equation}
that correspond to $pp$, $np$ and $nn$ configurations, respectively.  Since the 
transformation from $P^A_t$ to $\tilde{P}^{\tilde{A}}_t$ is a unitary rotation, 
the normalization is automatically correct:
\begin{equation}
 \Tr\left((\tilde{P}^{\tilde{A}}_t)^\dagger
 \tilde{P}^{\tilde{B}}_t\right)
 = \frac12 \delta_{\tilde{A}\tilde{B}} \,.
\end{equation}

In terms of these fields and projectors, the two-body interactions are
\begin{equation}
 \mathcal{L}_\mathrm{2d} = -d^{i\dagger}\left[\sigmad
 + \cd\left(\ii D_0+\frac{\vD^2}{4\MN}\right)\right]d^i
 + \yd\left[d^{i\dagger}\left(N^T P^i_d N\right)+\hc\right]
\label{eq:L-3S1}
\end{equation}
and
\begin{multline}
 \mathcal{L}_\mathrm{2t} = -t^{0\dagger}\left[\sigma_t
 + \ct\left(\ii D_0+\frac{\vD^2}{4\MN}\right)\right]t^0
 - t^{{-1}\dagger}\left[\sigmatpp
 + \ctpp\left(\ii D_0+\frac{\vD^2}{4\MN}\right)\right]t^{-1}\\
 - t^{{+1}\dagger}\left[\sigmatnn
 + \ctnn\left(\ii D_0+\frac{\vD^2}{4\MN}\right)\right]t^{+1}
 +\yt\left[t^{{\tilde A}\dagger}
 \left(N^T \tilde{P}^{\tilde A}_t N\right)+\hc\right] \, .
\label{eq:L-1S0}
\end{multline}
The covariant derivatives include the appropriate charge operators $\hat{Q}$. 
To keep the notation as simple as possible, we use the plain subscript ``$t$'' 
to refer to the $np$ channel from here on and use further qualifications only 
to denote $pp$ and $nn$ (the latter is not explicitly considered in the rest of 
this paper, except at the end of Sec. \ref{sec:Results}).  The parameters 
$\sigma_{d}$ and $\sigma_{t(,\cdot\cdot)}$ are related to the respective 
scattering lengths; each is actually a sum of contributions from various orders:
\begin{align}
 \sigmad^{\phantom{()}}
 &= \sigmad^{(0)} + \sigmad^{(1)} + \cdots \,, \\
 \sigma_{t(,\cdot\cdot)}^{\phantom{(01)}}
 &= \sigma_{t}^{(0)} + \sigma_{t(,\cdot\cdot)}^{(1)} + \cdots \,.
\label{eq:sigmas}
\end{align}
A more detailed discussion is given in Ref.~\cite{Konig:2014ufa}.  Departing 
from what is used there we follow here Ref.~\cite{Griesshammer:2004pe} and 
simply set
\begin{equation}
 \yd^2 = \yt^2 = \frac{4\pi}{\MN} \,.
\label{eq:simple-y}
\end{equation}
At the same time, the new parameters $\cd$ and $c_{t(,\cdot\cdot)}$ 
have been introduced to incorporate effective-range corrections, 
\cf~Fig.~\ref{fig:Corr-rd-rt}, starting at NLO:
\begin{align}
 \cd^{\phantom{(1)}} &= \cd^{(1)} + \cdots \,, \\
 c_{t(,\cdot\cdot)} &= c_{t}^{(1)} + \cdots \,.
\label{eq:cs}
\end{align}
In writing Eqs. \eqref{eq:sigmas} and \eqref{eq:cs} we imposed isospin symmetry
at the lowest order of each parameter. The reason is that isospin violation
comes from either electromagnetism or the up-down quark mass splitting.  These
are associated with mass scales $\alpha \MN \sim 7$ MeV and $m_u-m_d\sim 3$ MeV 
that are small compared to the breakdown scale $\LamNoPi$.

The three-nucleon interaction that is needed already at LO to 
renormalize the doublet-channel amplitude~\cite{Bedaque:1999ve} can be written 
as~\cite{Ando:2010wq,Griesshammer:2011md}
\begin{equation}
 \mathcal{L}_3=\frac{h}{3}N^\dagger\left[\yd^2\,
 d^{i\dagger} d^j \sigma^i \sigma^j+\yt^2\,t^{A\dagger} t^B \tau^A\tau^B
 - \yd\yt\left(d^{i\dagger} t^A \sigma^i \tau^A + \hc\right) \right]N \,,
\label{eq:L-3}
\end{equation}
where the coupling $h$ is also split up in various orders,
\begin{equation}
 h = h^{(0)} + h^{(1)} + \cdots \,.
\end{equation}
As we are going to show below, there is no need for an isospin-breaking
three-body force to NLO in our expansion.  Higher-order terms, including further 
isospin violation, are briefly discussed in Sec.~\ref{sec:Results}.

\section{Two-body sector}
\label{sec:TwoBody}

In this section we use the Lagrangian from the previous section to derive the 
two-body propagators and amplitudes which are basic ingredients of the 
three-body calculation described in the next section.

\subsection{Spin-triplet propagator}
\label{sec:SpinTriplet}

\begin{figure}[tb]
\centering
\includegraphics[clip]{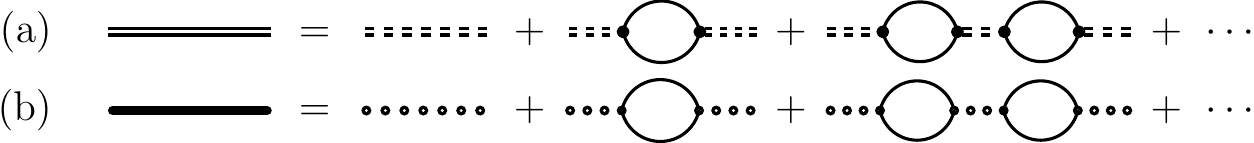}
\caption{Full dibaryon propagators in (a) the $^3S_1$ state (\ie, the deuteron)
and (b) the $^1S_0$ state. A single solid line represents the propagation
of a nucleon, a double dashed line a bare $d$ propagator, and a line of
circles a bare $t$ propagator.}
\label{fig:DibaryonProp}
\end{figure}

The dibaryon residual mass $\sigma_d$ represents the physics of the triplet 
$N\!N$ scattering length or, alternatively, the deuteron binding momentum 
$\gamma_d$.  For momenta $Q\sim \gamma_d\ll \LamNoPi$, the standard power 
counting of pionless EFT
applies~\cite{vanKolck:1997ut,Kaplan:1998tg,Kaplan:1998we,vanKolck:1998bw}.
The bare LO dibaryon propagator is simply $-\ii/\sigma_d^{(0)}$, and it has 
to be dressed by nucleon bubbles to all orders in order to get the full LO 
expression.  Summing up the geometric series shown diagrammatically in 
Fig.~\ref{fig:DibaryonProp}(a) gives
\begin{equation}
 \ii\Delta_{d}^{(0)}(p_0,\vp)
 = \frac{-\ii}{\sigma_{d}^{(0)} + y_{d}^2 I_0(p_0,\vp)} \,,
\end{equation}
with the bubble integral
\begin{multline}
 I_0(p_0,\vp) = \MN\int^\Lambda\ddq
 \frac{1}{\MN p_0 - \vp^2/4 - \vq^2 + \ii\eps} \\
 = {-}\frac{\MN}{4\pi}
 \left(\frac{2\Lambda}{\pi} - \sqrt{\frac{\vp^2}{4}-\MN p_0-\ii\eps}\right)
 +\OO(1/\Lambda) \,.
\label{eq:I0-cutoff}
\end{multline}
Here, we have used a simple momentum cutoff $\Lambda$ for regularization (not 
to be confused with the physical scale $\LamNoPi$ at which pionless EFT breaks 
down because new dynamical degrees of freedom are resolved).  To get 
expressions in the more commonly used power divergence subtraction (PDS) 
scheme~\cite{Kaplan:1998tg}, one has to replace $2\Lambda/\pi \rightarrow \mu$, 
where $\mu$ is a scale introduced through dimensional regularization.  Of 
course, in a properly renormalized theory the choice of regulator is arbitrary.
While power counting is clean and transparent in the PDS scheme, we find it 
convenient here to work with the simple cutoff instead, because this regulator 
is more straightforward when it comes to the consistent treatment of Coulomb 
contributions.  For completeness we note that Eq.~\eqref{eq:I0-cutoff} is valid 
up to corrections proportional to inverse powers of the cutoff, which we 
neglect here.

%%%%%%%%%%%%%%%%%%%%%%%%%%%%%%%%%%%%%%%%%%%%%%%%%%%%%%%%%%%%%%%%%%%%%%%%%%%%%%
%%%%%%%%%%%%%%%%%%%%%%%%%%%%%%%%%%%%%%%%%%%%%%%%%%%%%%%%%%%%%%%%%%%%%%%%%%%%%%

\begin{figure}[tb]
\centering
\begin{minipage}{0.49\textwidth}
 \centering
 \vcenteredhbox{\includegraphics[width=5.5em,clip]{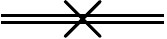}}
 \vcenteredhbox{\scalebox{1}{$\;\sim\;
 (-\ii\cd)\left(p_0-\frac{\vp^2}{4\MN}\right)$}}\\[0.77em]
 (a)
\end{minipage}\begin{minipage}{0.49\textwidth}
 \centering
 \vcenteredhbox{\includegraphics[width=5.5em,clip]{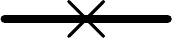}}
 \vcenteredhbox{\scalebox{1}{$\;\sim\;
 (-\ii\ct)\left(p_0-\frac{\vp^2}{4\MN}\right)$}}\\[0.77em]
 (b)
\end{minipage}
\caption{Dibaryon kinetic-energy corrections in (a) the $^3S_1$ state 
and (b) the $^1S_0$ state.}
\label{fig:Corr-rd-rt}
\end{figure}

Range corrections are accounted for by the dibaryon kinetic parameter $\cd$, 
which is included fully perturbatively here.  At NLO, we consider one insertion
of the operator shown in Fig.~\ref{fig:Corr-rd-rt}(a) between $\Delta_d^{(0)}$s.
Depending on the renormalization procedure, the insertion of $c_d^{(1)}$ might 
require a concomitant insertion of $\sigma_d^{(1)}$,
\begin{equation}
 \ii\Delta_d^{(1)}(p_0,\vp)
 = \ii\Delta_d^{(0)}(p_0,\vp)
 \times 
 \left[{-\ii}{\sigma_d^{(1)}}{-\ii}c_d^{(1)}
 \left(p_0-\frac{\vp^2}{4M_N}\right) \right]
 \times \ii\Delta_d^{(0)}(p_0,\vp) \,.
\label{eq:Delta-d-1}
\end{equation}

%%%%%%%%%%%%%%%%%%%%%%%%%%%%%%%%%%%%%%%%%%%%%%%%%%%%%%%%%%%%%%%%%%%%%%%%%%%%%%
%%%%%%%%%%%%%%%%%%%%%%%%%%%%%%%%%%%%%%%%%%%%%%%%%%%%%%%%%%%%%%%%%%%%%%%%%%%%%%

In the spin-triplet channel, we use the effective range expansion around the 
deuteron pole for renormalization and thus require that
\begin{equation}
 -\ii T(k)
 = (\ii y_d)^2\, \ii\Delta_{d}\!\left(p_0=k^2/\MN,\vp=\vZero\right)
 = \ii \frac{4\pi}{M_N}\frac{1}{k\cot\delta_{d}(k) - \ii k}\,,
\label{eq:Tnd}
\end{equation}
with the perturbative expansion
\begin{multline}
\frac{1}{k\cot\delta_{d}(k) - \ii k}
 = \frac{1}{\rule{0pt}{1.33em}{-}\gamd + \dfrac{\rd}2\big(k^2+\gamd^2\big)
 + \cdots - \ii k}
 = \frac{1}{{-}\gamd - \ii k}
 \left[1 + \frac{\rd}{2}\frac{k^2+\gamd^2}{\gamd+\ii k}
 + \cdots\right] \, ,
\label{eq:Delta-d-renorm}
\end{multline}
where $\gamd = \sqrt{\mathstrut\MN E_d}\simeq 45.7$ MeV~\cite{vanderLeun:1982aa} 
is the deuteron binding momentum and $\rho_d\simeq 1.765$ 
fm~\cite{deSwart:1995ui} the deuteron effective range.

At LO, the effective range $\rho_d\sim 1/\LamNoPi$ does not contribute and 
one simply has
\begin{equation}
 \sigmad^{(0)} = \frac{2\Lambda}{\pi}-\gamd \,.
\label{eq:sigmad0}
\end{equation}
The corresponding propagator $\Delta_d^{(0)}$ is given by the term outside the 
parentheses in Eq.~\eqref{eq:Delta-d-renorm}, up to an additional minus sign. 
 
The corrections in the parentheses come from subleading orders.  The first 
of these are $\OO(Q/\LamNoPi, \gamma_d/\LamNoPi)$, or NLO, and 
\begin{equation}
 \sigmad^{(1)} = \frac{\rd\gamd^2}{2} \mathtext{,}
 \cd^{(1)} = \frac{\MN\rd}{2} \,.
\label{eq:sigmad1}
\end{equation}

\subsection{Spin-singlet propagators}
\label{sec:SpinSinglet}

In the spin-singlet channel, in the absence of Coulomb effects, we can follow 
the same procedure as in the spin-triplet channel (see 
Figs.~\ref{fig:DibaryonProp}(b) and~\ref{fig:Corr-rd-rt}(b)), and arrive at the 
LO propagator
\begin{equation}
 \ii\Delta_{t}^{(0)}(p_0,\vp)
 = \frac{-\ii}{\sigma_{t}^{(0)} + y_{t}^2 I_0(p_0,\vp)} \,.
\end{equation}
At NLO, we have
\begin{equation}
 \ii\Delta_{t}^{(1)}(p_0,\vp)
 = \ii\Delta_{t}^{(0)}(p_0,\vp)\times
 \bigg[{-}\ii\sigma_{t}^{(1)}
 - \ii c_{t}^{(1)} \left( p_0-\frac{\vp^2}{4M_N}\right) \bigg]
 \times \ii\Delta_{t}^{(0)}(p_0,\vp) \,.
\label{eq:Delta-t-NLO-1}
\end{equation}
Renormalization is then performed by matching $\Delta_t$ to the ${}^1S_0$ 
effective range expansion around zero momentum,
\begin{equation}
 k\,\cot\delta_t(k) = {-}\frac{1}{a_t} + \frac{\rnt}{2} k^2 + \OO(k^4) \,,
\label{eq:ERE-1S0}
\end{equation}
where $a_t\simeq -23.7~\fm$ and $\rnt\approx2.73~\fm$~\cite{Preston:1975} are
the $np$ scattering length and effective range, respectively.  As for the 
triplet, $r_t\sim 1/\LamNoPi$, but $|a_t| \gg 1/\gamma_d$.

\subsubsection{Standard approach}
\label{sec:SpinSinglet-Conventional}

Typically, one is interested in low momenta $Q\sim 1/|a_t|\ll \LamNoPi$ and
demands that at leading order the scattering length is reproduced.  
Effective-range and higher corrections are included perturbatively, since they 
are $\OO(Q/\LamNoPi, 1/(|a_t|\LamNoPi))$.  Up to \NLO in this 
{\it standard} scheme one sets
\begin{equation}
 \sigma_t^{(0,\mathrm{st})} = \frac{2\Lambda}{\pi} - \frac{1}{a_t}
 \mathtext{,}
 \sigma_t^{(1,\mathrm{st})} = 0
 \mathtext{,}
 \ct^{(1,\mathrm{st})} = \frac{\MN\rnt}{2} \,.
\label{eq:tparconv}
\end{equation}
There is no adjustment of $\sigma_t$ when the effective-range expansion is 
performed around the zero-energy threshold.

\subsubsection{Unitarity limit}
\label{sec:SpinSinglet-Unitarity}

The standard approach makes sense for momenta $Q\sim 1/|a_t|$, so at LO we 
have contributions from both the unitarity cut and the scattering length.
However, if we are interested in momenta $1/|a_t|\ll Q\ll \LamNoPi$, we are 
close to the unitarity limit and take instead
\begin{equation}
 \sigma_t^{(0)} = \frac{2\Lambda}{\pi}
 \mathtext{,}
 \sigma_t^{(1)} = {-}\frac{1}{a_t}
 \mathtext{,}
 \ct^{(1)} = \frac{\MN\rnt}{2} \,,
\label{eq:sigma-t-unitarity}
\end{equation}
so the actual finiteness of the scattering length only enters as a perturbative 
correction.  The difference with respect to Eq.~\eqref{eq:tparconv} is a 
result of $|a_t| \gg 1/\gamma_d$.  For example, for $Q\sim \gamma_d$, we are 
performing an extra expansion in the ratio $1/(|a_t|\gamma_d)\ll 1$.  Meanwhile, 
other effective range parameters require no special treatment, being similar in 
both channels---for example, $r_t \sim \rho_d$.

Of course, to the extent that this expansion works one might as well resum 
$\sigma_t^{(1)}$ and use Eq.~\eqref{eq:tparconv} with 
$\sigma_t^{(0,\mathrm{st})} = \sigma_t^{(0)} + \sigma_t^{(1)}$
over the whole range $Q\ll \LamNoPi$.  However, there are also advantages in 
keeping a strict ordering.  It makes clear, for example, that the singlet LO is 
parameter free.  Also, as we show shortly, it matches well with the 
perturbative expansion of electromagnetic effects.

\subsection{Coulomb insertions}
\label{sec:Coulomb}

Simple dimensional analysis shows that the Coulomb expansion is in powers of 
$\alpha M_N/Q$, while other electromagnetic corrections are suppressed by
at least $(Q/\LamNoPi)^2$.  To NLO we need only keep the contribution from 
static Coulomb photons.  In this case, the matching discussed above is correct 
for the $np$ part of the ${}^1S_0$ dibaryon.  If one neglects strong isospin 
breaking, it can also be used to describe the $nn$ component.  For $pp$ 
configurations, however, one has to use the Coulomb-modified effective 
range expansion~\cite{Bethe:1949yr,Kong:1999sf},
\begin{equation}
 C_\eta^2 \left(k\cot\delta_{t,pp}(k) - \ii k\right) + \alpha\MN H(\eta)
 = {-}\frac{1}{\app} + \frac{\rpp}{2} k^2 + \cdots
\label{eq:ERE-pp}
\end{equation}
because electromagnetic effects dominate the very-low-energy scattering
regime, $Q\lsim\alpha\MN$.  These are encoded in the Coulomb parameter
\begin{equation}
 \eta(k) = \frac{\alpha\MN}{2k} \,,
\label{eq:eta-k}
\end{equation}
the Gamow factor
\begin{equation}
 C_\eta^2 = \frac{2\pi\eta}{\ee^{2\pi\eta} - 1} \,, 
\end{equation}
and the function
\begin{equation}
 H(\eta)= \psi({\ii}\eta) +\frac{1}{2{\ii}\eta} - \log({\ii}\eta) \,,
\label{eq:H-eta}
\end{equation}
with $\psi$ the derivative of the Euler Gamma function.  Here $\app \simeq 
-7.81~\fm$ and $r_C\simeq 2.79~\fm$~\cite{Bergervoet:1988zz} are the 
Coulomb-modified scattering length and effective range, respectively.  One 
arrives at Eq.~\eqref{eq:ERE-pp} after subtracting the pure Coulomb amplitude 
with Coulomb phase shifts
\begin{equation}
 \exp(2\ii\sigma_0) = \Gamma(1+\ii\eta)/\Gamma(1-\ii\eta)
\end{equation}
from the full amplitude.

\subsubsection{Kong+Ravndal approach}
\label{sec:Coulomb-KR}

Since $\alpha M_N\sim 1/|a_t|$, describing physics at the scale of the $^1S_0$ 
virtual state requires also a resummation of Coulomb exchange.  This has been 
studied in detail by Kong and Ravndal in Ref.~\cite{Kong:1999sf} in a setup 
without dibaryon fields.  We refer to that reference for details and note here 
that the relation of our parameter $\sigma_{t,pp}^{(0,\mathrm{st})}$ to the 
$C_0$ of Kong and Ravndal is $\sigma_{t,pp}^{(0,\mathrm{st})} = {-}{4\pi}/{(\MN 
C_0)}$.

\begin{figure}[tb]
\centering
\includegraphics[clip]{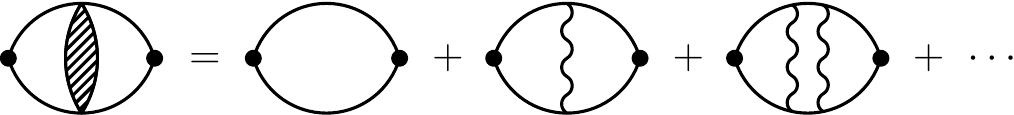}
\caption{Fully dressed proton--proton bubble. A wavy line denotes 
a Coulomb photon exchange.}
\label{fig:DressedBubble}
\end{figure}

\medskip
The key ingredient is the fully dressed Coulomb bubble shown in
Fig.~\ref{fig:DressedBubble}.  From an evaluation of the Coulomb Green's
function, Kong and Ravndal find that it is given by the divergent bubble
integral
\begin{equation}
 J_0(k) = \MN \int \ddq\,\frac{2\pi\eta(q)}{\ee^{2\pi\eta(q)}-1}
 \frac{1}{k^2-q^2+\ii\eps} \,,
\label{eq:J0-int}
\end{equation}
and they separate the divergent part with a subtraction at $k=0$:
\begin{spliteq}
 J_0(k) &= \MN \int \ddq\,\frac{2\pi\eta(q)}{\ee^{2\pi\eta(q)}-1}
 \frac{k^2}{q^2(k^2-q^2+\ii\eps)}
 - \MN \int \ddq\,\frac{2\pi\eta(q)}{\ee^{2\pi\eta(q)}-1}\frac{1}{q^2} \\
 &\equiv J_0^{\text{fin}}(k) + J_0^{\text{div}} \,,
\label{eq:J0-fin-div}
\end{spliteq}
where the finite piece is
\begin{equation}
 J_0^{\text{fin}}(k) = {-}\frac{\alpha\MN^2}{4\pi} H(\eta) \,.
\label{eq:J0-fin}
\end{equation}
With a simple momentum cutoff, the divergent part is
\begin{equation}
 J_0^{\text{div}} = {-}\frac{\MN\Lambda}{2\pi^2}
 + \frac{\alpha\MN^2}{4\pi}
 \left(\log\frac{2\Lambda}{\alpha\MN}
 - \EulerGamma\right) + \OO(1/\Lambda) \,,
\label{eq:J0-div-Cutoff}
\end{equation}
where $\EulerGamma\approx0.5772$ is the Euler--Mascheroni constant, and 
we neglect higher-order corrections in $1/\Lambda$.

%%%%%%%%%%%%%%%%%%%%%%%%%%%%%%%%%%%%%%%%%%%%%%%%%%%%%%%%%%%%%%%%%%%%%%%%%%%%%%
%%%%%%%%%%%%%%%%%%%%%%%%%%%%%%%%%%%%%%%%%%%%%%%%%%%%%%%%%%%%%%%%%%%%%%%%%%%%%%

Resumming now this dressed bubble along with $\sigma_{t,pp}^{(0,\mathrm{st})}$,
the singlet dibaryon propagator in the $pp$ channel becomes
\begin{equation}
 \ii\Delta_{t,pp}^{(0,\mathrm{st})}(p_0,\vp)
 = \dfrac{{-}\ii}
 {\rule{0pt}{1.66em}\sigma_{t,pp}^{(0,\mathrm{st})} - \dfrac{2\Lambda}{\pi}
 + \alpha\MN\left(\log\dfrac{2\Lambda}{\alpha\MN}- \EulerGamma\right)
 - \alpha\MN H(\eta)} \,,
\label{eq:Delta-t-pp-KR}
\end{equation}
with
\begin{equation}
 \eta = \eta\!\left(\ii\sqrt{\vp^2/4 - \MN p_0 - \ii\eps}\right) \,.
\label{eq:eta-gen}
\end{equation}
Matching to the Coulomb-modified effective range expansion is performed via
the part of the T-matrix where Coulomb interferes with the short-range
interactions,
\begin{equation}
 -\ii T_{SC}(k) = C_\eta^2 \ee^{2\ii\sigma_0}
 (\ii y_t)^2 \, \ii \Delta_{t,pp}\!\left(p_0=k^2/\MN,\vp=\vZero\right) 
 = \ii \frac{4\pi}{M_N}
 \frac{\ee^{2\ii\sigma_0}}{k\cot\delta_{t,pp}(k) - \ii k} \,.
\label{eq:coulmatch}
\end{equation}
The additional factors here compared to the $np$ component, Eq.~\eqref{eq:Tnd},
arise from the inclusion of initial and final-state Coulomb interactions in 
order to get the amplitude from the propagator.  Combining this relation with 
Eq.~\eqref{eq:ERE-pp} one finds that both the Gamow factor and the pure Coulomb 
phase shift drop out and one arrives at
\begin{equation}
 \sigma_{t,pp}^{(0,\mathrm{st})} = -\frac{1}{\app} + \dfrac{2\Lambda}{\pi}
 - \alpha\MN\left(\log\dfrac{2\Lambda}{\alpha\MN}- \EulerGamma\right) \,.
\label{eq:renorm-sigma-t-pp}
\end{equation}
Range corrections were also considered in Ref.~\cite{Kong:1999sf}.

\subsubsection{Separate leading-order resummation}
\label{sec:Coulomb-Unitarity}

Here we develop a new approach that allows us to consider a fully 
isospin-symmetric leading order in the spin-singlet channel, including the 
$pp$ part.  Since $\app$ is still large compared to the typical nuclear length 
scale set by the inverse pion mass ($1/m_\pi \sim 1.4~\fm$), in the momentum 
window $\alpha M_N \lsim 1/\app \ll Q \ll \LamNoPi$ we remain close to 
unitarity in the singlet channels.  We should be able to treat Coulomb 
perturbatively along with finite scattering-length corrections.  The key idea 
is that with the new counting, the LO singlet propagator now behaves exactly 
like $1/k$, \ie, it has the same infrared behavior as Coulomb contributions, 
for which the relevant parameter (in the $pp$ system) is
$\eta = \alpha\MN/(2k)$.

\begin{figure}[tb]
\centering
\begin{minipage}{0.333\textwidth}
 \centering
 \vcenteredhbox{\includegraphics[width=6em,clip]{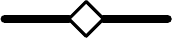}}
 \vcenteredhbox{\scalebox{1.2}{$\;\sim\;{-}\ii\sigma_{t,pp}^{(1)}$}}\\[0.77em]
 (a)
\end{minipage}
\begin{minipage}{0.333\textwidth}
 \centering
 \includegraphics[width=6.35em,clip]{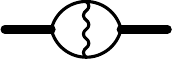}\\[0.52em]
 (b)
\end{minipage}
\caption{Corrections in the $pp$ channel: (a) Coulomb-corrected 
scattering length; (b) one-photon exchange.}
\label{fig:Corr-at-bub}
\end{figure}

At LO we thus have, with an isospin-symmetric $\sigma_t^{(0)}$ satisfying 
Eq.~\eqref{eq:sigma-t-unitarity},
\begin{equation}
 \ii\Delta_{t,pp}^{(0)}(p_0,\vp) = \ii\Delta_t^{(0)}(p_0,\vp)
 = \frac{{-}\ii}
 {\rule{0pt}{1.66em}\sqrt{\frac{\vp^2}{4}-\MN p_0-\ii\eps}} \,.
\label{eq:Delta-t-pp-LO}
\end{equation}
At NLO, we need perturbative insertions not only of $\sigma_{t,pp}^{(1)}$ 
(Fig.~\ref{fig:Corr-at-bub}(a)) and $c_{t}^{(1)}$ 
(Fig.~\ref{fig:Corr-rd-rt}(b)), but also of a single-photon exchange, see 
Fig.~\ref{fig:Corr-at-bub}(b).  Using the notation of Ref.~\cite{Kong:1999sf}, 
we call the single-photon piece $\delta I_0(k)$ and find for the correction to 
the $pp$ propagator:
\begin{multline}
 \ii\Delta_{t,pp}^{(1)}(p_0,\vp)
 = \ii\Delta_{t,pp}^{(0)}(p_0,\vp)\times
 \bigg[{-}\ii\sigma_{t,pp}^{(1)}
 - \ii c_{t}^{(1)} \left( p_0-\frac{\vp^2}{4M_N}\right) \\
 - \ii \yt^2\delta I_0\!\left(\ii\sqrt{\vp^2/4-\MN p_0-\ii\eps}\right)\bigg]
 \times \ii\Delta_{t,pp}^{(0)}(p_0,\vp) \,.
\label{eq:Delta-pp-NLO-1}
\end{multline}
In order to match to the Coulomb-modified effective range expansion, we work in 
the $pp$ center-of-mass frame in the remainder of this section.  From the 
expression for $\delta I_0(k)$ in Appendix~\ref{sec:SinglePhotonBubble}, we find 
for $p_0=k^2/\MN$, $\vp=\vZero$, and $\eta$ as defined in Eq.~\eqref{eq:eta-k} 
that
\begin{equation}
 \delta I_0(k) = \frac{\alpha\MN^2}{4\pi}
 \left[\log\ii\eta
 + \log\frac{2\Lambda}{\alpha M_N} - C_\zeta\right] + \OO(1/\Lambda) \,,
\label{eq:delta-I0}
\end{equation}
where $C_\zeta \simeq 1.119$ and we again drop terms proportional to 
inverse powers of $\Lambda$.  The log divergence is absorbed into 
$\sigma_{t,pp}^{(1)}$, whereas the $\log(\ii\eta)$ constitutes the one-photon 
contribution to $H(\eta)$ as defined in Eq.~\eqref{eq:H-eta}.

%%%%%%%%%%%%%%%%%%%%%%%%%%%%%%%%%%%%%%%%%%%%%%%%%%%%%%%%%%%%%%%%%%%%%%%%%%%%%%
%%%%%%%%%%%%%%%%%%%%%%%%%%%%%%%%%%%%%%%%%%%%%%%%%%%%%%%%%%%%%%%%%%%%%%%%%%%%%%

In order to find the renormalization condition for $\sigma_{t,pp}^{(1)}$,
we now consider smaller $Q$, which requires the resummation of both 
$\sigma_{t,pp}^{(1)}$ and Coulomb exchange.  We determine $\sigma_{t,pp}^{(1)}$
from this calculation, and then use it in the regime where Coulomb is 
perturbative, to order $\alpha$.  For this, it is important to be consistent 
about which finite terms get absorbed into $\sigma_{t,pp}^{(1)}$ along with the 
logarithmic divergence.

\begin{figure}[tb]
\centering
\includegraphics[clip]{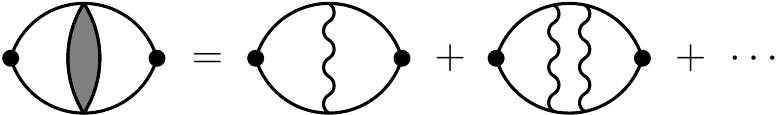}
\caption{Dressed proton--proton bubble excluding the empty piece.}
\label{fig:DressedBubble-new}
\end{figure}

The resummation of Coulomb produces a new dressed bubble, shown in 
Fig.~\ref{fig:DressedBubble-new}.  The important point is that this new dressed 
bubble excludes the empty piece (no photon exchange inside the bubble) because 
that has already been resummed at LO.  We denote by $\delta J_0(k)$ the part 
remaining after subtracting the single-photon piece $\delta I_0(k)$.  
In Appendix~\ref{sec:FullyDressedBubble-T} it is demonstrated that
\begin{equation}
 \delta J_0(k) = {-}\frac{\alpha\MN^2}{4\pi}
 \left[\psi(\ii\eta) + \frac{1}{2\ii\eta} + C_\Delta\right]
 + \frac{\MN}{4\pi} \ii k \,,
\label{eq:delta-J0}
\end{equation}
where $C_\Delta \approx 0.579$.  Hence, we find for the new dressed bubble
\begin{equation}
 \delta I_0(k) + \delta J_0(k)
 = \frac{\alpha\MN^2}{4\pi}
 \bigg[\log\frac{2\Lambda}{\alpha\MN} - C_\zeta - C_\Delta - H(\eta)\bigg]
 + \frac{\MN}{4\pi} \ii k \,.
\label{eq:delta-I0-J0}
\end{equation}

%%%%%%%%%%%%%%%%%%%%%%%%%%%%%%%%%%%%%%%%%%%%%%%%%%%%%%%%%%%%%%%%%%%%%%%%%%%%%%
%%%%%%%%%%%%%%%%%%%%%%%%%%%%%%%%%%%%%%%%%%%%%%%%%%%%%%%%%%%%%%%%%%%%%%%%%%%%%%

The resummation of $\sigma_{t,pp}^{(1)}$ and the new dressed bubble is 
now straightforward.  We consider here explicitly the on-shell 
case in the center-of-mass frame---$p_0=k^2/\MN$, $\vp=\vZero$---and find
\begin{spliteq}
 \ii\Delta_{t,pp}^{\text{res}}(k)
 &= \ii\Delta_{t,pp}^{(0)}(k) + \ii\Delta_{t,pp}^{(0)}(k)
    \left({-}\ii\sigma_{t,pp}^{(1)}
    - \ii \yt^2\delta I_0(k) - \ii \yt^2\delta J_0(k)\right)
    \ii\Delta_{t,pp}^{(0)}(k)
    + \cdots \\
 &= \frac{{-}\ii}{\rule{0pt}{1.66em}
 \underbrace{\sigma_t^{(0)} - \dfrac{2\Lambda}{\pi}}_{\null=0}
 \null + \sigma_{t,pp}^{(1)}
 + \alpha\MN\left(\log\dfrac{2\Lambda}{\alpha\MN} - C_\zeta - C_\Delta\right)
 - \alpha\MN H(\eta)} \,,
\label{eq:Delta-pp-res}
\end{spliteq}
where the explicit imaginary part $\ii k$ cancels.  Now, from 
Eqs.~\eqref{eq:coulmatch} and~\eqref{eq:ERE-pp}, 
\begin{equation}
 \sigma_{t,pp}^{(1)} = {-}\frac{1}{a_C}
 - \alpha\MN\left(\log\dfrac{2\Lambda}{\alpha\MN}
 - C_\zeta - C_\Delta\right) \,,
%\qquad c_{t}^{(1)} = \frac{\MN r_C}{2} \,.
\label{eq:renorm-sigma-t-pp-1}
\end{equation}
Our renormalization of $\sigma_{t,pp}^{(0)}+\sigma_{t,pp}^{(1)}$ is consistent
with Eq.~\eqref{eq:renorm-sigma-t-pp}, but with a different constant finite 
piece.  As discussed in more detail in the Appendix, the reason for this 
change is that we have isolated here the divergent piece---the bubble with a 
single photon exchange---and only regularized that, whereas 
Ref.~\cite{Kong:1999sf} regularizes the fully resummed bubble---including the 
empty part---at once.  Effectively, this amounts to using different 
regularization schemes.  Our approach has the advantage that it allows for a 
consistent matching between the perturbative and nonperturbative regimes.

The expression for the $pp$ phase shift in the regime $\alpha M_N \lsim 1/|\app|
\ll Q \ll \LamNoPi$ can now be found by inserting Eqs.~\eqref{eq:Delta-t-pp-LO} 
and~\eqref{eq:Delta-pp-NLO-1} into Eq.~\eqref{eq:coulmatch}:
\begin{equation}
 k\cot\delta_{t,pp}(k)
 = {-}\frac{1}{\app} + \alpha M_N C_\Delta+\frac{r_t}{2} k^2 
 + \alpha\MN \log\left(\frac{\alpha\MN}{2k}\right) + \cdots \,.
\label{eq:ERE-pppert}
\end{equation}
This is consistent with a direct expansion of Eq.~\eqref{eq:ERE-pp} in powers of 
$\alpha M_N/Q$ and $1/(|\app|Q)$.  Using 
\begin{equation}
 \psi(i\eta) = \frac{i}{\eta} -C_E + \OO(\eta) \,,
\label{eq:psiexpanded}
\end{equation}
one finds Eq.~\eqref{eq:ERE-pppert} with $C_\Delta=C_E$ and
\begin{equation}
 r_C^{\text{NLO}} = r_t \,.
\label{eq:rCprediction}
\end{equation}
Moreover, the same result as in Eq.~\eqref{eq:ERE-pppert} can be obtained in a 
direct calculation that includes all $\OO(\alpha)$ diagrams contributing to $pp$ 
scattering, treated in perturbation theory. 

As discussed in Appendix~\ref{sec:FullyDressedBubble-T}, a numerical calculation 
indeed yields $C_\Delta\approx0.579$, very close to $\EulerGamma\approx0.5772$.
Equation \eqref{eq:ERE-pppert} gives circumstantial evidence that in fact 
$C_\Delta=\EulerGamma$, and we expect that increased numerical accuracy would 
reveal the same.  Lacking a formal proof, however, we keep $C_\Delta$ in 
the expressions here and merely note that the deviation from $\EulerGamma$ is 
negligible for the results presented later.

Equation~\eqref{eq:ERE-pppert} is in agreement with the Nijmegen phase-shift 
analysis~\cite{Stoks:1993tb} up to about 15\% for $k \gsim 60~\MeV$.  In 
particular, it captures the correct slope at intermediate momenta, which is a 
consequence of the fact that Eq.~\eqref{eq:rCprediction} works at the 3\% 
level.  The assumption of isospin symmetry in $\ct^{(1)}$, which means that we 
can neglect the splitting in the spin-singlet ranges, is a good one.

\section{Three-body sector}
\label{sec:ThreeBody}

We now consider the three-body sector in order to calculate the \Triton and
\He binding energies.  These bound states arise in the $nd$ and $pd$ 
spin-doublet channels, respectively.  Since we do not reorganize the pionless 
EFT expansion in the $N\!N$ spin-triplet channel, no changes are needed in the 
$nd$ and $pd$ spin-quartet channels with respect to
Refs.~\cite{Bedaque:1997qi,Bedaque:1998mb,Bedaque:1999vb,Gabbiani:1999yv,
Vanasse:2013sda,Konig:2013cia}.

We take the point of view that the scale that characterizes these bound states, 
the triton binding momentum $\gamma_T$, is comparable to the deuteron binding 
momentum $\gamma_d$, but both are much larger than $\alpha M_N$, $1/|a_t|$, and 
$1/|\app|$.  Thus the bound-state energies can be expanded not only in powers 
of $Q/\LamNoPi$ but also of $\aleph_0/Q$, where $\aleph_0 \sim \alpha\MN \sim 
1/|a_t| \sim 1/|\app|$.\footnote{Note that $\aleph_0$ is an extension of 
the original definition~\cite{vanKolck:1997ut,vanKolck:1998bw} to include 
additional scales, in particular $\alpha M_N$.}  For simplicity we pair the two 
expansions by taking $Q\sim (\aleph_0 \LamNoPi)^{1/2}$.

\subsection{More formalism}
\label{sec:ThreeBody-Formalism}

According to the standard power 
counting~\cite{Bedaque:1997qi,Bedaque:1998mb,Bedaque:1999vb,Gabbiani:1999yv}
one-nucleon exchange between nucleon and dibaryon has to be treated exactly.  
To NLO it is sufficient to consider the $S$-wave projected one-nucleon-exchange 
diagram at energy $E$,
\begin{equation}
 \KS(E;k,p) \equiv \frac{1}{kp}\;
 Q_0\left(\frac{k^2+p^2-\MN E-\ii\eps}{kp}\right)
 = \frac{1}{2kp}\log\!\left(
 \frac{k^2+p^2+kp-\MN E-\ii\eps}{k^2+p^2-kp-\MN E-\ii\eps}\right) \,,
\label{eq:KS}
\end{equation}
where $k$ ($p$) is the incoming (outgoing) center-of-mass momentum.  It is well
known~\cite{Bedaque:1999ve,Hammer:2000nf,Hammer:2001gh,Bedaque:2002yg,
Afnan:2003bs,Griesshammer:2005ga} that at LO the resummation of one-nucleon 
exchange is renormalized by the three-body force given in Eq.~\eqref{eq:L-3}, 
with
\begin{equation}
 h^{(0)}= \frac{\MN H(\Lambda)}{\Lambda^2} \,,
\end{equation}
where $\Lambda$ is now the momentum cutoff applied in the three-body equations
discussed below, and $H(\Lambda)$ a known log-periodic function of the cutoff
that depends on a three-body parameter $\Lambda_*$.  Here we follow the 
procedure employed in Ref.~\cite{Bedaque:1997qi} and much of the subsequent 
literature, in which the two-body cutoff is taken to be very large and 
$\OO(1/\Lambda)$ terms in Eqs.~\eqref{eq:I0-cutoff} and~\eqref{eq:delta-I0}
are neglected.\footnote{This is effectively equivalent to using dimensional 
regularization in order to renormalize the two-body sector first.}

In order to calculate bound-state energies and perturbative corrections to 
them, we use the formalism and notation of 
Refs.~\cite{Konig:2014ufa,Koenig:2013} to study the $Nd$ doublet channel.  We 
introduce a three-component vector of vertex functions in channel space,
\begin{equation}
 \vec{\BS}\equiv\left(\BSda,\BSdb,\BSdc\right)^T \,,
\end{equation}
where $\BSda$ corresponds to the deuteron channel and $\BSdb$ and $\BSdc$ are
the $np$ and $pp$/$nn$ components of the ${}^1S_0$ multiplet, respectively. 
These are given by (properly normalized~\cite{Konig:2011yq}) solutions of the 
homogeneous equation
\begin{equation}
 \vec{\BS} = (\hat{K}\hat{D})\otimes\vec{\BS} \mathtext{,} E = \EB \,,
\label{eq:BS-IntEq}
\end{equation}
where $\otimes$ represents an integral over the intermediate momentum, 
$\hat{D}$ is a diagonal matrix of dibaryon propagators written in the form
\begin{equation}
 D_{d,t}(E;q) = \Delta_{d,t}\!\left(E-\frac{q^2}{2\MN};q\right) \,,
\end{equation}
and
\begin{equation}
 \hat{K}\equiv\begin{pmatrix}
 -g_{dd}\left(\KS+\frac{2H(\Lambda)}{\Lambda^2}\right) &
 g_{dt}\left(3\KS+\frac{2H(\Lambda)}{\Lambda^2}\right) &
 g_{dt}\left(3\KS+\frac{2H(\Lambda)}{\Lambda^2}\right)\\[\skrowspace]
 g_{dt}\left(\KS+\frac{2H(\Lambda)}{3\Lambda^2}\right) &
 g_{tt}\left(\KS-\frac{2H(\Lambda)}{3\Lambda^2}\right) &
 -g_{tt}\left(\KS+\frac{2H(\Lambda)}{3\Lambda^2}\right)\\[\skrowspace]
 g_{dt}\left(2\KS+\frac{4H(\Lambda)}{3\Lambda^2}\right) &
 -g_{tt}\left(2\KS+\frac{4H(\Lambda)}{3\Lambda^2}\right) &
 -g_{tt}
%\times
\frac{4H(\Lambda)}{3\Lambda^2}
 \end{pmatrix} \,.
\label{eq:K-B}
\end{equation}
%and
%
The factors $g_{dd}$ {\it etc.}~contain the coupling constants $\yd$ and $\yt$. 
In our present conventions, we simply have
\begin{equation}
 g_{dd} = g_{tt} = g_{dt} = 2\pi \,.
\end{equation}
For illustration, we note that with a single channel and no three-body force, 
the homogeneous integral equation written out explicitly would be
\begin{equation}
 \BS(p) = \frac{1}{\pi}\int_0^\Lambda\dd q\,q^2\,\KS(E;p,q)\,
 \Delta\!\left(E-\frac{q^2}{2\MN};q\right) \BS(q) \,,
\end{equation}
where $\Delta$ denotes a generic dibaryon propagator, and $\Lambda$ is the 
momentum cutoff employed in the three-body sector.

To study doublet-channel $nd$ scattering, we have to solve the 
Lippmann--Schwinger equation.  This can be done in a simpler two-channel 
formalism, where it becomes
\begin{multline}
 \begin{pmatrix}\TSda \\[\skrowspace] \TSdb\end{pmatrix}
  = \begin{pmatrix}
  g_{dd}\left(\KS+\frac{2H(\Lambda)}{\Lambda^2}\right)\\[\skrowspace]
  -g_{dt}\left(3\KS+\frac{2H(\Lambda)}{\Lambda^2}\right)
 \end{pmatrix} \\
 +\begin{pmatrix}
 -g_{dd}D_d\left(\KS+\frac{2H(\Lambda)}{\Lambda^2}\right) &
 g_{dt}D_t\left(3\KS+\frac{2H(\Lambda)}{\Lambda^2}\right) \\[\skrowspace]
 g_{dt}D_d\left(3\KS+\frac{2H(\Lambda)}{\Lambda^2}\right) &
 -g_{tt}D_t\left(\KS+\frac{2H(\Lambda)}{\Lambda^2}\right)
 \end{pmatrix}
 \otimes\begin{pmatrix}\TSda\\[\skrowspace]\TSdb\end{pmatrix} \,.
\label{eq:IntEq-TSdb}
\end{multline}
Again, the notation is the same as in Refs.~\cite{Konig:2014ufa,Koenig:2013}.  
The $nd$ phase shift is then obtained from the on-shell amplitude,
\begin{equation}
 \delta_{\text{$n$--$d$}}(k) = \frac{1}{2\ii}
 \log\!\left(1+\frac{2\ii k\MN}{3\pi} Z_0\TSda(E_k;k,k)\right)
 \mathtext{,} E_k = \frac{3k^2}{4\MN}-\frac{\gamd^2}{\MN} \,,
\label{eq:delta-nd}
\end{equation}
where $Z_0$ is the deuteron wavefunction renormalization determined by
\begin{equation}
 Z_0^{-1} = \ii\frac{\partial}{\partial p_0}
 \left.\frac{1}{\Delta_d(p)}\right|_{p_0 =-\frac{\gamd^2}{\MN},\,\vp=0} \,.
\label{eq:Z0}
\end{equation}
The scattering length is determined by the on-shell amplitude in the 
limit $k\to0$,
\begin{equation}
 \andDoublet = {-}\frac{\MN}{3\pi}\lim\nolimits_{k\to0} Z_0\TSda(E_k;k,k) \,.
\label{eq:and-2}
\end{equation}

\subsection{Perturbative range corrections}
\label{sec:ThreeBody-RangeCorrections}

\begin{figure}[tb]
\centering
\begin{minipage}{0.49\textwidth}
 \centering
 \includegraphics[width=0.4275\textwidth,clip]{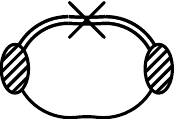}\\[0.77em]
 (a)
\end{minipage}\begin{minipage}{0.49\textwidth}
\centering
 \vspace*{0.5em}
 \includegraphics[width=0.441\textwidth,clip]{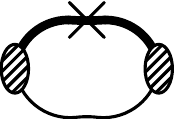}\\[0.77em]
 (b)
\end{minipage}
\caption{Range corrections contributing to the trinucleon binding energy
in perturbation theory.  A shaded oval represents a vertex function.}
\label{fig:DeltaE-rdt}
\end{figure}

Before we discuss Coulomb matrix elements between vertex functions, it is
instructive to consider effective-range corrections in this framework.  As we 
emphasized in Sec. \ref{sec:TwoBody}, range corrections are $\OO(Q/\LamNoPi)$.
In analogy to Eq.~\eqref{eq:Delta-d-renorm}, we write
\begin{multline}
 D_d(E;q) = D_d^{(0)}(E;q) + D_d^{(1)}(E;q) + \cdots \\
 = \frac{{-}1}{-\gamd+\sqrt{3q^2/4-\MN E-\ii\eps}}
 \times\left[1 + \frac{\rd}{2}\frac{\left(3q^2/4-\MN E-\gamd^2\right)}
 {-\gamd+\sqrt{\strut3q^2/4-\MN E-\ii\eps}}
 + \cdots \right] \,,
\label{eq:Prop-d-expansion}
\end{multline}
with an analogous expression for the spin-singlet part.  Suppose now we have 
solved the homogeneous equation~\eqref{eq:BS-IntEq} at leading order.  The 
binding-energy shift due to the deuteron effective range, shown in 
Fig.~\ref{fig:DeltaE-rdt}(a), is then given by
\begin{equation}
 \Delta E^{(1)}_{\rd} = \frac{\rd}{4\pi^2}\int_0^\Lambda \dd q \,q^2\,
 \abs{\BSda(q)}^2 \frac{\left(3q^2/4-\MN E-\gamd^2\right)}{\rule{0pt}{1.8em}
 \bigg({-}\gamd+\sqrt{\strut3q^2/4-\MN E-\ii\eps}\bigg)^{\!2}} \,.
\label{eq:DeltaE-rd}
\end{equation}
More details about this can be found in 
Refs.~\cite{Vanasse:2013sda,Vanasse:2014kxa}.  In particular, we note that the 
correction $h^{(1)}$ to the three-body force is fitted to keep whatever 
physical parameter has been used at LO (triton binding energy or 
doublet-channel scattering length) unchanged.

%%%%%%%%%%%%%%%%%%%%%%%%%%%%%%%%%%%%%%%%%%%%%%%%%%%%%%%%%%%%%%%%%%%%%%%%%%%%%%
%%%%%%%%%%%%%%%%%%%%%%%%%%%%%%%%%%%%%%%%%%%%%%%%%%%%%%%%%%%%%%%%%%%%%%%%%%%%%%

In the {\it standard} counting used previously (finite scattering lengths at 
LO), there is a contribution analogous to Eq.~\eqref{eq:DeltaE-rd} with 
$\gamd\to1/a_t$ in the denominator (and no $\gamd$ in the numerator).  In our 
new scheme with the spin-singlet LO in the unitarity limit, the range 
corrections in the spin-singlet channels are particularly simple.  For example, 
from Fig.~\ref{fig:DeltaE-rdt}(b) we have
\begin{equation}
 \Delta E^{(1)}_{\rnt,\text{b1}} = \frac{3\rnt}{4\pi^2}
 \int_0^\Lambda \dd q \,q^2\,\abs{\BSdb(q)}^2 \,,
\end{equation}
and an analogous expression that involves $\BSdc$.  NLO corrections that are 
linear in the range have been known for a long time to generate cutoff 
dependence that can be compensated by 
$h^{(1)}$~\cite{Bedaque:1999ve,Hammer:2001gh,Griesshammer:2005ga}.  For three 
bosons at unitarity, this divergence was discussed in
Refs.~\cite{Platter:2008cx,Ji:2010su}.  At the same time, we have of course now 
perturbative insertions of the scattering length, \eg,
\begin{equation}
 \Delta E^{(1)}_{a_t,\text{b1}} = \frac{3}{2\pi^2a_t}\int_0^\Lambda \dd q\,q^2\,
 \frac{\abs{\BSdb(q)}^2}{3q^2/4+\MN\EBp} \,,
\label{eq:3bscattlencor}
\end{equation}
which are corrections of $\OO(\aleph_0/Q)$ relative to LO.

\subsection{Coulomb matrix elements}
\label{sec:ThreeBody-Coulomb}

Range corrections as discussed in the previous section apply in general
to both $nd$ and $pd$ systems.  In the bound-state regime, they simply 
correspond to matrix elements between trinucleon wavefunctions that are
diagonal in momentum space (only one loop integral is required to calculate 
them) as well as in cluster-configuration (channel) 
space~\cite{Griesshammer:2004pe}.

\begin{figure}[tb]
\centering
\begin{minipage}{0.333\textwidth}
 \centering
 \includegraphics[width=0.63\textwidth,clip]{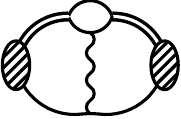}\\[0.77em]
 (a)
\end{minipage}\begin{minipage}{0.333\textwidth}
 \centering
 \vspace*{0.47em}
 \includegraphics[width=0.71\textwidth,clip]{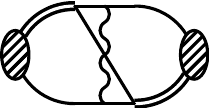}\\[0.77em]
 (b)
\end{minipage}\begin{minipage}{0.333\textwidth}
\centering
 \vspace*{0.47em}
 \includegraphics[width=0.71\textwidth,clip]{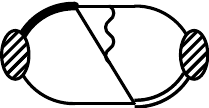}\\[0.77em]
 (c)
\end{minipage}
\caption{Convergent diagrams contributing to the $^3\mathrm{H}$--$^3\mathrm{He}$
binding energy difference in perturbation theory.}
\label{fig:DeltaE-conv}
\end{figure}

\begin{figure}[tb]
\centering
\begin{minipage}{0.49\textwidth}
 \centering
 \vspace*{0.25em}
 \includegraphics[width=0.4275\textwidth]{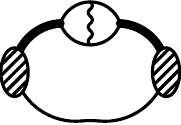}\\[0.77em]
 (a)
\end{minipage}\begin{minipage}{0.49\textwidth}
\centering
 \vspace*{0.5em}
 \includegraphics[width=0.441\textwidth]{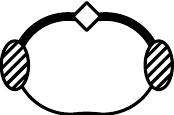}\\[0.77em]
 (b)
\end{minipage}
\caption{Divergent Coulomb bubble diagram (a) contributing to the \Triton--\He 
binding energy difference along with the associated counterterm (b).}
\label{fig:DeltaE-bub-C}
\end{figure}

Now we want to include Coulomb corrections, which are $\OO(\aleph_0/Q)$.
In general, contributions to the \Triton--\He  energy splitting $\Delta E$ can 
be non-diagonal in both spaces.  We get such contributions when we calculate
$\Delta E$ in perturbation theory, which comes from diagram topologies shown 
in Figs.~\ref{fig:DeltaE-conv} and~\ref{fig:DeltaE-bub-C}.  This approach 
starts by taking the trinucleon state to be the triton in a three-channel 
formalism.  This way, one can easily isolate the channel that corresponds to the 
$pp$ configuration in \He.  Such a calculation was carried out in 
Ref.~\cite{Konig:2014ufa} for the diagrams in Fig.~\ref{fig:DeltaE-conv}, which 
give convergent results.  We briefly summarize this calculation here before 
getting to the new diagrams in Fig.~\ref{fig:DeltaE-bub-C}.

%%%%%%%%%%%%%%%%%%%%%%%%%%%%%%%%%%%%%%%%%%%%%%%%%%%%%%%%%%%%%%%%%%%%%%%%%%%%%%
%%%%%%%%%%%%%%%%%%%%%%%%%%%%%%%%%%%%%%%%%%%%%%%%%%%%%%%%%%%%%%%%%%%%%%%%%%%%%%

%%%%%%%%%%%%%%%%%%%%%%%%%%%%%%%%%%%%%%%%%%%%%%%%%%%%%%%%%%%%%%%%%%%%%%%%%%%%%%
%%%%%%%%%%%%%%%%%%%%%%%%%%%%%%%%%%%%%%%%%%%%%%%%%%%%%%%%%%%%%%%%%%%%%%%%%%%%%%

The diagram shown in Fig.~\ref{fig:DeltaE-conv}(a) is still diagonal in 
cluster-configuration space.  In order to calculate it, we need the kernel 
function corresponding to the photon being exchanged between the deuteron 
bubble and the individual proton.  It is given by~\cite{Konig:2011yq}
\begin{equation}
 K_\mathrm{bubble}(E;k,p)
 = {-}\alpha\MN \int_{-1}^1\dd\!\cos\theta\,
 \frac{\mathcal{I}_\mathrm{bubble}(E;\vk,\vp)}{(\vk-\vp)^{2}+\lambda^2}
 \mathtext{,} \vk\cdot\vp = kp\cos\theta \,,
\label{eq:K-bubble}
\end{equation}
with
\begin{equation}
 \mathcal{I}_\mathrm{bubble}(E;\vk,\vp)
 = \frac{\arctan\left(\frac{2\vp^2-\vk^2-\vk\cdot\vp}
 {\sqrt{3\vk^2-4\MN E-\ii\eps}\sqrt{(\vk-\vp)^2}}\right)
 +\arctan\left(\frac{2\vk^2-\vp^2-\vk\cdot\vp }
 {\sqrt{3\vp^2-4\MN E-\ii\eps}\sqrt{(\vk-\vp)^2}}\right)}
 {\sqrt{(\vk-\vp)^2}}
\label{eq:I-bubble-pd} \,,
\end{equation}
and where $\lambda$ is a photon mass introduced for regularization in the
infrared.  In practice, we do not perform the angular integral numerically but
rather use the explicit $S$-wave projection given in 
Ref.~\cite{Vanasse:2014kxa}.  The contribution to the energy shift is then 
given by
\begin{multline}
 \Delta E^{(1)}_{\ref{fig:DeltaE-conv}(a)} = \frac{1}{2\pi^3}
 \int_0^\Lambda \dd q_1 \,q_1^2 \int_0^\Lambda \dd q_2 \,q_2^2\,
 \BSda(q_1)\,D_d^{(0)}(\EB,q_1) \\
 \times K_\mathrm{bubble}(E;q_1,q_2)
 \,D_d^{(0)}(\EB,q_2)\,\BSda(q_2) \,.
\label{eq:DeltaE-bub}
\end{multline}
The analogous diagram with $np$ singlet propagators (not shown explicitly
in Fig.~\ref{fig:DeltaE-conv}) is given by essentially the same expression with
the replacements $\BSda \rightarrow \BSdb$ and $D_d \rightarrow D_t$.  For the 
``box'' and ``triangle'' contributions, Figs.~\ref{fig:DeltaE-conv}(b) and~(c), 
the corresponding kernel functions are~\cite{Konig:2014ufa}
\begin{multline}
 K_{\text{box}}(E;k,p) = -\alpha\MN \\
 \times\frac12\int_{-1}^1\dd\!\cos\theta\,
 \Bigg\{\frac{\arctan\Big(\frac{2\vp^2-\vk^2-\vk\cdot\vp}
 {\sqrt{3\vk^2-4\MN E-\ii\eps}\sqrt{(\vk-\vp)^2}}\Big)
 +\arctan\Big(\frac{2\vk^2-\vp^2-\vk\cdot\vp }
 {\sqrt{3\vp^2-4\MN E-\ii\eps}\sqrt{(\vk-\vp)^2}}\Big)}
 {(\vk^2+\vp^2+\vk\cdot\vp-\MN E-\ii\eps)\sqrt{(\vk-\vp)^2}} \\
 - \frac{\lambda}{(\vk^2+\vp^2+\vk\cdot\vp-\MN E-\ii\eps)^2}
 + \OO(\lambda^2) \Bigg\} \,,
\label{eq:K-box}
\end{multline}
and
\begin{subequations}%
\begin{equation}
 K_\text{tri}^{(\text{out})}(E;k,p) = -\alpha\MN \\
 \times\frac12\int_{-1}^1\dd\!\cos\theta
 \frac{\mathcal{I}_{\text{tri}}(E;\vk,\vp)}
 {\vk^2+\vp^2+\vk\cdot\vp-\MN E-\ii\eps} \,,
\label{eq:K-tri-out}
\end{equation}
\begin{equation}
 K_\text{tri}^{(\text{in})}(E;k,p) = K_\text{tri}^{(\text{out})}(E;p,k) \,,
\label{eq:K-tri-in}
\end{equation}
\label{eq:K-tri}%
\end{subequations}%
where the superscripts ``out'' and ``in'' indicate whether the Coulomb-photon 
exchange is on the left or right side of the diagram.  This notation is 
taken over from Ref.~\cite{Konig:2014ufa}, which allowed for incoming and 
outgoing $pd$ states.  The loop function appearing in Eq.~\eqref{eq:K-tri-out} 
is given by
\begin{multline} 
 \mathcal{I}_{\text{tri}}(E;\vk,\vp)
 = \frac{\ii}{2\sqrt{\vk^2/4+\vk\cdot\vp+\vp^2}} \\
 \times\Bigg\{
 \log\left(\frac{\ii(\vk^2/2-\vk\cdot\vp-\vp^2-\lambda^2-\MN E-\ii\eps)}
 {\sqrt{\vk^2/4+\vk\cdot\vp+\vp^2}}
 +2\sqrt{\lambda^2+3\vk^2/4-\MN E-\ii\eps}\right) \\
 -\log\left(\frac{\ii(\vk^2+\vp^2+\vk\cdot\vp-\lambda^2-\MN E-\ii\eps)}
 {\sqrt{\vk^2/4+\vk\cdot\vp+\vp^2}}+2\lambda\right)\Bigg\} \,.
\end{multline}
As for $K_\mathrm{bubble}(E;k,p)$, explicit $S$-wave projections where the 
integral over $\cos\theta$ has been carried out analytically can be found in 
Ref.~\cite{Vanasse:2014kxa}.  Resulting contributions to the energy shift are 
of the form
\begin{multline}
 \Delta E^{(1)}_{\ref{fig:DeltaE-conv}(b)} = \frac{1}{2\pi^3}
 \int_0^\Lambda \dd q_1 \,q_1^2 \int_0^\Lambda \dd q_2 \,q_2^2\,
 \BSda(q_1)\,D_d^{(0)}(\EB,q_1) \\
 \times K_\mathrm{box}(E;q_1,q_2)
 \,D_d^{(0)}(\EB,q_2)\,\BSda(q_2)
\label{eq:DeltaE-box}
\end{multline}
and
\begin{multline}
 \Delta E^{(1)}_{\ref{fig:DeltaE-conv}(c)} = {-}\frac{3}{2\pi^3}
 \int_0^\Lambda \dd q_1 \,q_1^2 \int_0^\Lambda \dd q_2 \,q_2^2\,
 \BSdb(q_1)\,D_{t}^{(0)}(\EB,q_1) \\
 \times K_\mathrm{tri}^{(\mathrm{out})}(E;q_1,q_2)
 \,D_d^{(0)}(\EB,q_2)\,\BSda(q_2) \,,
\label{eq:DeltaE-tri}
\end{multline}
with analogous expressions for equivalent topologies but different combinations 
of dibaryon propagators and vertex functions (see Ref.~\cite{Konig:2014ufa} for 
details).

The above summarizes the diagrams included in the perturbative calculation of 
Ref.~\cite{Konig:2014ufa}; all these contributions are convergent as the cutoff 
$\Lambda$ is increased.  As mentioned in the introduction, the contribution 
from the diagram shown in Fig.~\ref{fig:DeltaE-bub-C}(a) has not been included 
so far.  This diagram is logarithmically divergent, but this is precisely the 
same divergence of the one-photon bubble that we isolated in the new treatment 
of the two-body sector discussed in Sec.~\ref{sec:Coulomb-Unitarity}.  Hence, 
it can be renormalized by including it together with the counterterm diagram 
shown in Fig.~\ref{fig:DeltaE-bub-C}(b), which is proportional to 
$\sigma_{t,pp}^{(1)}$ as given in Eq.~\eqref{eq:renorm-sigma-t-pp-1}.  The 
resulting contribution to the energy shift, written out explicitly, is
\begin{multline}
 \Delta E^{(1)}_{\text{\ref{fig:DeltaE-bub-C}(a+b)}}
 = \frac{3}{4\pi^2} \int_0^\Lambda \dd q \,q^2\,
 \frac{\abs{\BSdc(q)}^2}{\strut3q^2/4 + \MN\EBp} \\
 \times \left\{\dfrac{1}{\app} - \alpha\MN\left[C_\Delta
 + \log\!\left(\dfrac{\alpha\MN}{2\sqrt{\mathstrut \MN E_B+3q^2/4}}\right)
 \right]\right\} \,.
\label{eq:DeltaE-aC}
\end{multline}
Note that the constant $C_\zeta$ drops out here against the same contribution 
from the photon bubble, \cf~Eq.~\eqref{eq:delta-I0}.  Expanding the 
\emph{renormalized} $pp$ propagator of Refs.~\cite{Ando:2010wq,Konig:2014ufa} in 
$\alpha$ gives Eq.~\eqref{eq:DeltaE-aC} with $C_\Delta \to \EulerGamma$,
as expected from their similar values (\cf the discussion in 
Sec.~\ref{sec:Coulomb-Unitarity}).

We stress here that the new approach takes all spin-singlet propagators in the 
unitarity limit at LO.  In the $pp$ channel, the finite scattering length 
$\app$ is included together with the single-photon bubble contribution, 
resulting in Eq.~\eqref{eq:DeltaE-aC}.  At the same time, we also include linear 
insertions of $1/a_t$ in the $np$ spin-singlet channel, as given in 
Eq. \eqref{eq:3bscattlencor}.

\section{Results and discussion}
\label{sec:Results}

We summarize our new expansion as follows: at leading order, we include
\begin{itemize}
\item the standard $N\!N$ spin-triplet (pionless) amplitude (parameter $\gamd$),
\item the unitary $N\!N$ spin-singlet amplitude (parameter-free),
\item a contact three-body force (parameter $\Lambda_*$). 
\end{itemize}
Our new NLO includes\footnote{These NLO contributions induce corrections to the 
spin-triplet two- and three-body force parameters that already appeared at LO, 
but these corrections introduce no new parameters.}
\begin{itemize}
\item the effective range in the $N\!N$ spin-triplet channel (parameter 
$\rho_d$),
\item the isospin-symmetric range in the $N\!N$ spin-singlet channel (parameter 
$r_t$), 
\item a scattering-length correction to unitarity in the $N\!N$ spin-singlet 
$np$ and $nn$ channels (parameter $a_t$),
\item a scattering-length correction to unitarity in the $N\!N$ $pp$ channel
(parameter $\app$), 
\item one-photon exchange (parameter $\alpha= 1/137$).
\end{itemize}
The two-body parameters we use in our numerical calculation are summarized
in Table~\ref{tab:Params}.  For the nucleon mass we take $\MN = 938.918~\MeV$.

%%%%%%%%%%%%%%%%%%%%%%%%%%%%%%%%%%%%%%%%%%%%%%%%%%%%%%%%%%%%%%%%%%%%%%%%%%%%%%
\begin{table}[tb]
\centering
 \begin{tabular}{ccc}
  Parameter & Value & Ref.\\
  \hline\hline
  \rule{0pt}{1.1em}$\gamd$ & $45.7~\MeV$ & \cite{vanderLeun:1982aa} \\
  $\rd$ & $1.765~\fm$ & \cite{deSwart:1995ui} \\
  $a_t$ & $-23.714~\fm$ & \cite{Preston:1975} \\
  $\rnt$ & $2.73~\fm$ & \cite{Preston:1975} \\
  $a_C$ & $-7.8063~\fm$ & \cite{Bergervoet:1988zz}
 \end{tabular}
\caption{Parameters used for the numerical calculation.}
\label{tab:Params}
\end{table}
%%%%%%%%%%%%%%%%%%%%%%%%%%%%%%%%%%%%%%%%%%%%%%%%%%%%%%%%%%%%%%%%%%%%%%%%%%%%%%

Unitarity in the $N\!N$ spin singlet at LO means that our results for the 
binding energies and scattering of nuclei differ from previous calculations, 
for example the \Triton binding energy and $nd$ scattering in the doublet 
channel~\cite{Bedaque:1999ve,Hammer:2000nf,Hammer:2001gh,Bedaque:2002yg,
Afnan:2003bs}.  In order to facilitate the comparison with existing, 
standard-LO results, we also consider below an ``incomplete'' new NLO, where
we set $\rho_d = r_t = 0$.

\begin{figure}[tb]
\centering
\includegraphics[width=0.7\textwidth,clip]{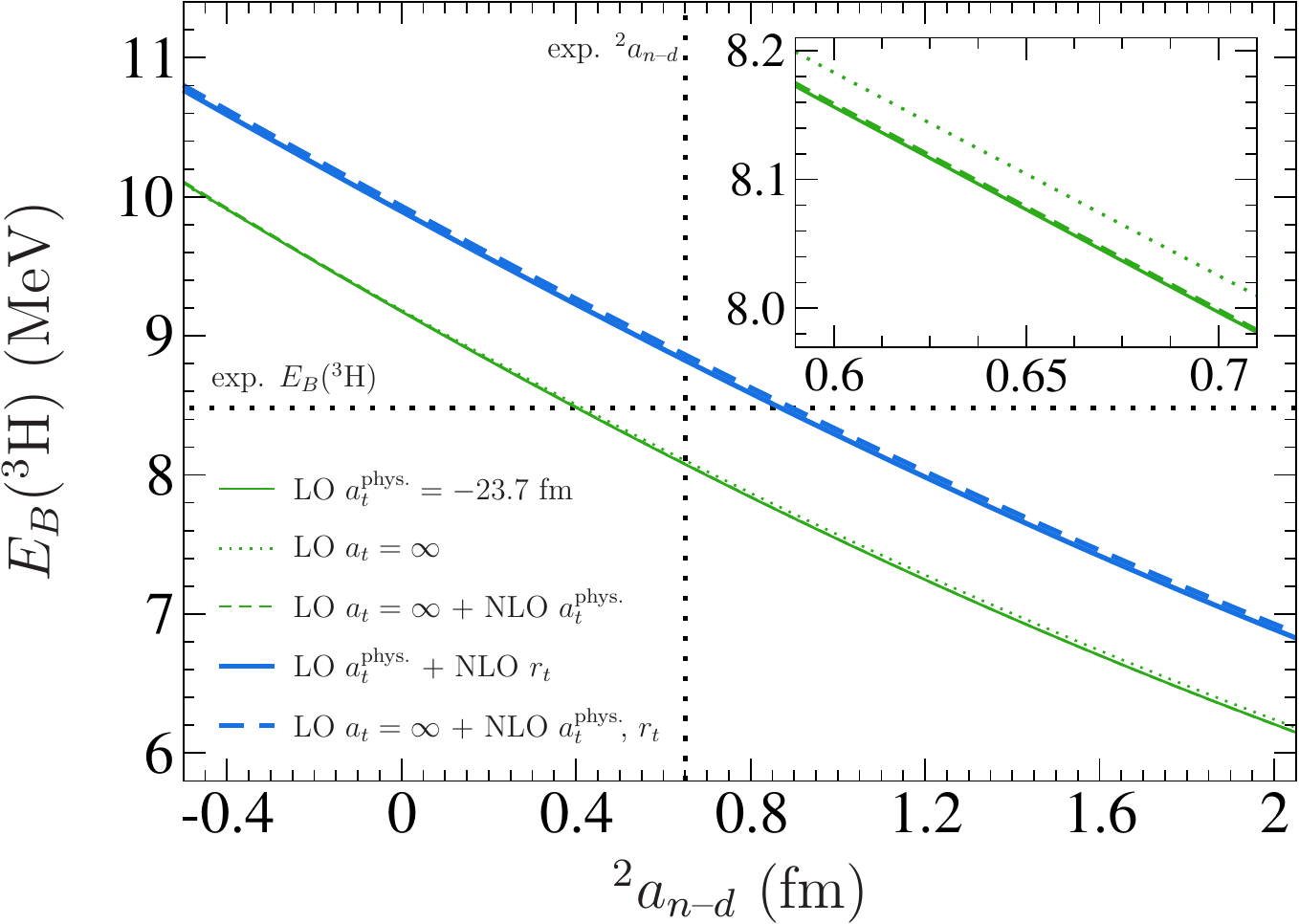}
\caption{Correlation (Phillips line) between \Triton binding energy (in \MeV)
and doublet $nd$ scattering length (in \fm) at LO and NLO.  The light (green) 
and dark (blue) solid lines are the results of the standard expansion at LO and 
LO+NLO.  The light (green) dotted line is the new LO, the light (green) dashed 
line is the new LO+NLO with ranges set to zero, and the dark (blue) dashed line 
is the full new LO+NLO.  All curves are for a three-body cutoff 
$\Lambda=2.4$~GeV.  Horizontal and vertical (black) dotted lines indicate 
experimental values for binding energy and scattering length, respectively.}
\label{fig:Phillips-U}
\end{figure}

In the spin singlet we perform an extra $\aleph_0/Q$ expansion on top of the 
standard $Q/\LamNoPi$ expansion.  We first show that the $\aleph_0/Q$ expansion 
gives results that are in good agreement with the standard leading-order $N\!N$ 
spin-singlet amplitude in the absence of Coulomb effects.  As $\Lambda_*$ is 
varied with fixed $N\!N$ input, doublet-channel observables change in a 
correlated way.  The simplest example is the Phillips 
line~\cite{Phillips:1968zze} in the plane of \Triton binding energy and $nd$ 
scattering length, see Fig.~\ref{fig:Phillips-U}.  Five curves are shown for a 
three-body cutoff $\Lambda=2.4$~GeV; effects from further increasing the cutoff 
are negligible.  In three of the curves the ranges are set to zero.  We see that 
the new LO curve (with $a_t\to \infty$) is within 1\% of the standard LO curve 
(with $a_t$ at its physical value), and the new LO+(incomplete)NLO (with 
$a_t\to \infty$ at LO and $1/a_t$ at its physical value treated in first-order 
perturbation theory, but zero range) is closer still.  The inset magnifies a 
region of the plot to show the small differences among these curves.  This 
agreement is not fortuitous and survives the inclusion of range corrections,
displayed in the other two curves.  Again, the new LO+NLO curve ($a_t\to \infty$ 
at LO, and both physical $1/a_t$ and ranges in first-order perturbation theory) 
is very close to the standard LO+NLO ($a_t$ at its physical value at LO, ranges 
in first-order perturbation theory).

%%%%%%%%%%%%%%%%%%%%%%%%%%%%%%%%%%%%%%%%%%%%%%%%%%%%%%%%%%%%%%%%%%%%%%%%%%%%%%
%%%%%%%%%%%%%%%%%%%%%%%%%%%%%%%%%%%%%%%%%%%%%%%%%%%%%%%%%%%%%%%%%%%%%%%%%%%%%%

In Fig.~\ref{fig:Phillips-U} we also indicate the experimental values of the 
\Triton binding energy and doublet $nd$ scattering length by, respectively, 
horizontal and vertical lines.  Leading-order curves (new as well as standard) 
lie close to the experimental point.  Next-order curves (new as well as 
standard) are small shifts in the direction of data, overshooting a bit.

\begin{figure}[tb]
\centering
\includegraphics[width=0.7\textwidth,clip]{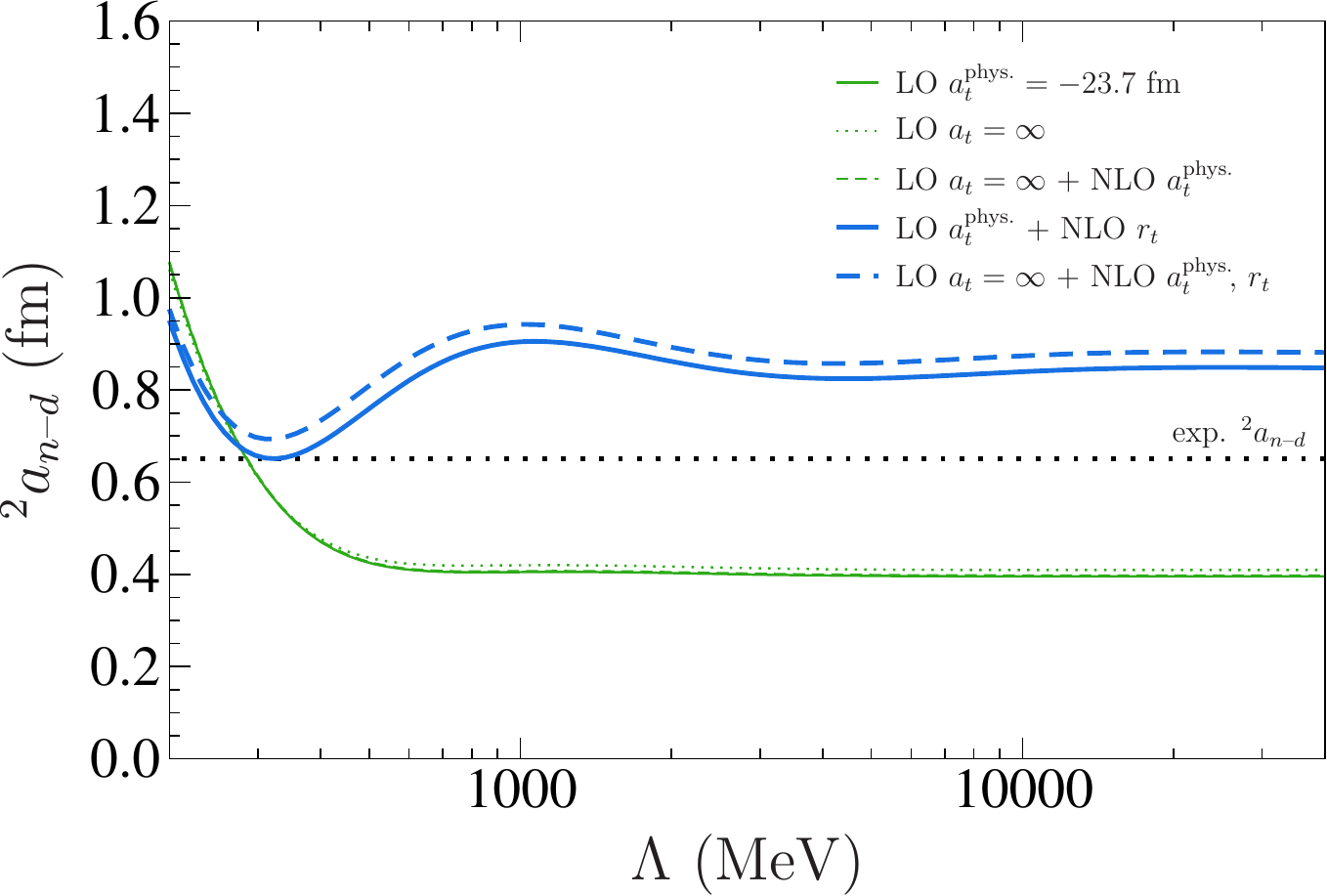}
\caption{$nd$ doublet-channel scattering length (in $\fm$) as a function of the 
cutoff $\Lambda$ (in $\MeV$).  Numerically, the limit in Eq.~\eqref{eq:and-2} 
has been taken by setting $k=0.01~\MeV$.  Notation as in 
Fig.~\ref{fig:Phillips-U}.}
\label{fig:2and-U}
\end{figure}

\begin{figure}[tb]
\centering
\includegraphics[width=0.7\textwidth,clip]{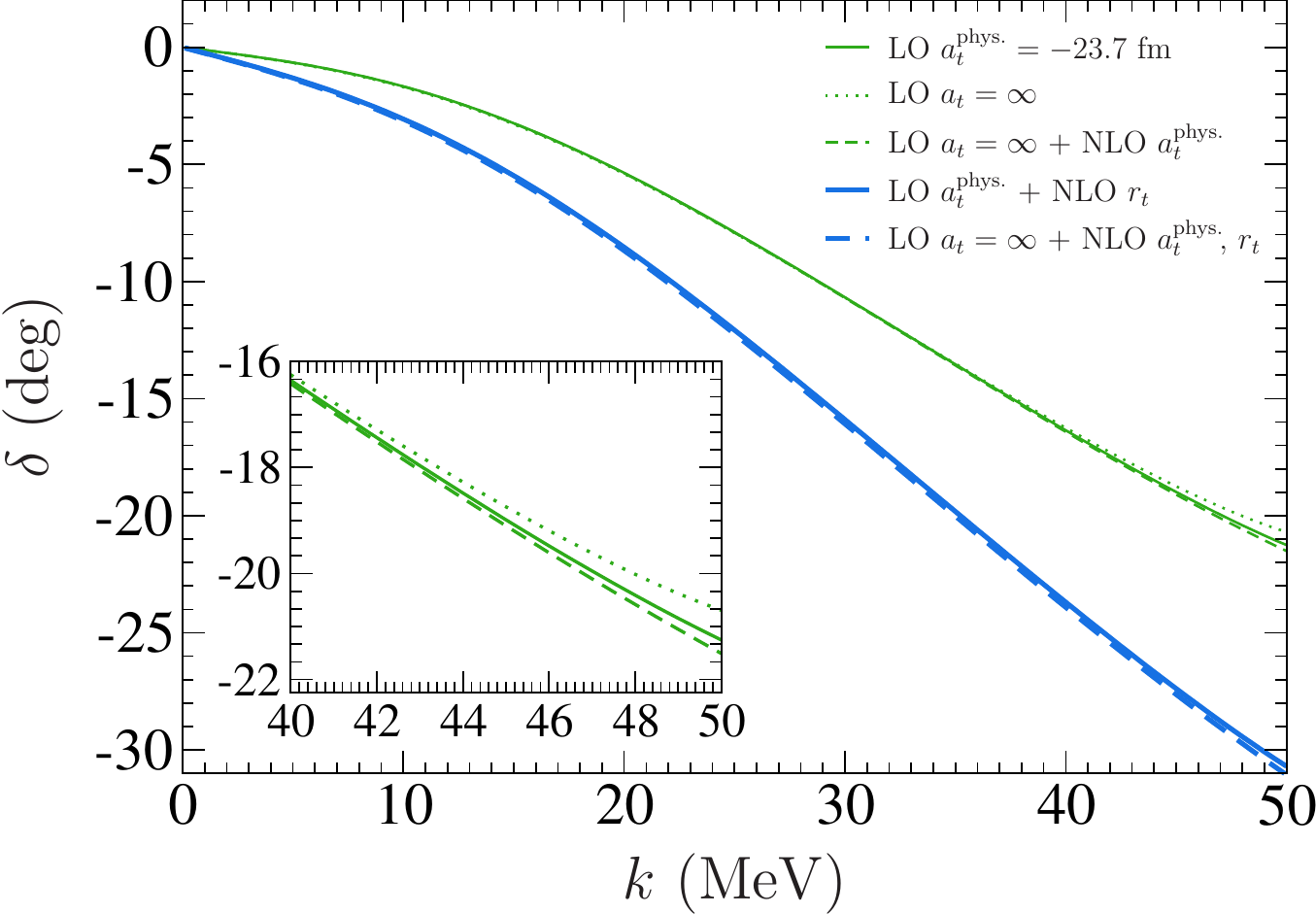}
\caption{$nd$ spin-doublet phase shift (in degrees) at LO and NLO
as function of the center-of-mass momentum (in $\MeV$).  Notation as in
Fig.~\ref{fig:Phillips-U}.}
\label{fig:Phase-D-U}
\end{figure}

We can use either the binding energy or the doublet-channel scattering length to
determine $\Lambda_*$, and then the other is a prediction that nearly agrees 
with data.  Here we use the \Triton binding energy, $\EBp(\Triton)=8.48~\MeV$, 
as input.  At LO this is done by adjusting $h^{(0)}$; at NLO, $h^{(1)}$ ensures 
that the \Triton binding energy remains at its experimental value.  The same 
procedure is used in the standard expansion with $a_t$ in the LO propagators, 
and our values for $h^{(0,1)}$ come out very close to results in that 
approach~\cite{Vanasse:2014kxa}.  Then the $nd$ scattering length converges as 
the cutoff $\Lambda$ increases, as shown in Fig.~\ref{fig:2and-U}, where 
five curves analogous to those in Fig.~\ref{fig:Phillips-U} are displayed.  
More generally, Fig.~\ref{fig:Phase-D-U} shows the predictions for the 
doublet-channel $nd$ phase shifts at low momenta.  Again, the effects of 
treating the finite value of the singlet scattering length in perturbation 
theory are small.

%%%%%%%%%%%%%%%%%%%%%%%%%%%%%%%%%%%%%%%%%%%%%%%%%%%%%%%%%%%%%%%%%%%%%%%%%%%%%%

%%%%%%%%%%%%%%%%%%%%%%%%%%%%%%%%%%%%%%%%%%%%%%%%%%%%%%%%%%%%%%%%%%%%%%%%%%%%%%
%%%%%%%%%%%%%%%%%%%%%%%%%%%%%%%%%%%%%%%%%%%%%%%%%%%%%%%%%%%%%%%%%%%%%%%%%%%%%%

Thus, the $\aleph_0/Q$ expansion works quite well in the absence of Coulomb 
interactions.  In fact, range corrections seem larger than those from the finite 
singlet scattering length, which suggests that the $\aleph_0/Q$ expansion works 
better than the $Q/\LamNoPi$ expansion.  With the \Triton channel properly 
renormalized and the three-body force fixed, we now consider Coulomb corrections 
to the \He binding energy.  Because our LO Lagrangian is isospin-symmetric, the 
\Triton--\He binding energy difference vanishes in our new LO, but it is a 
prediction---a low-energy theorem---at NLO.

\begin{figure}[tb]
\centering
\includegraphics[width=0.7\textwidth,clip]{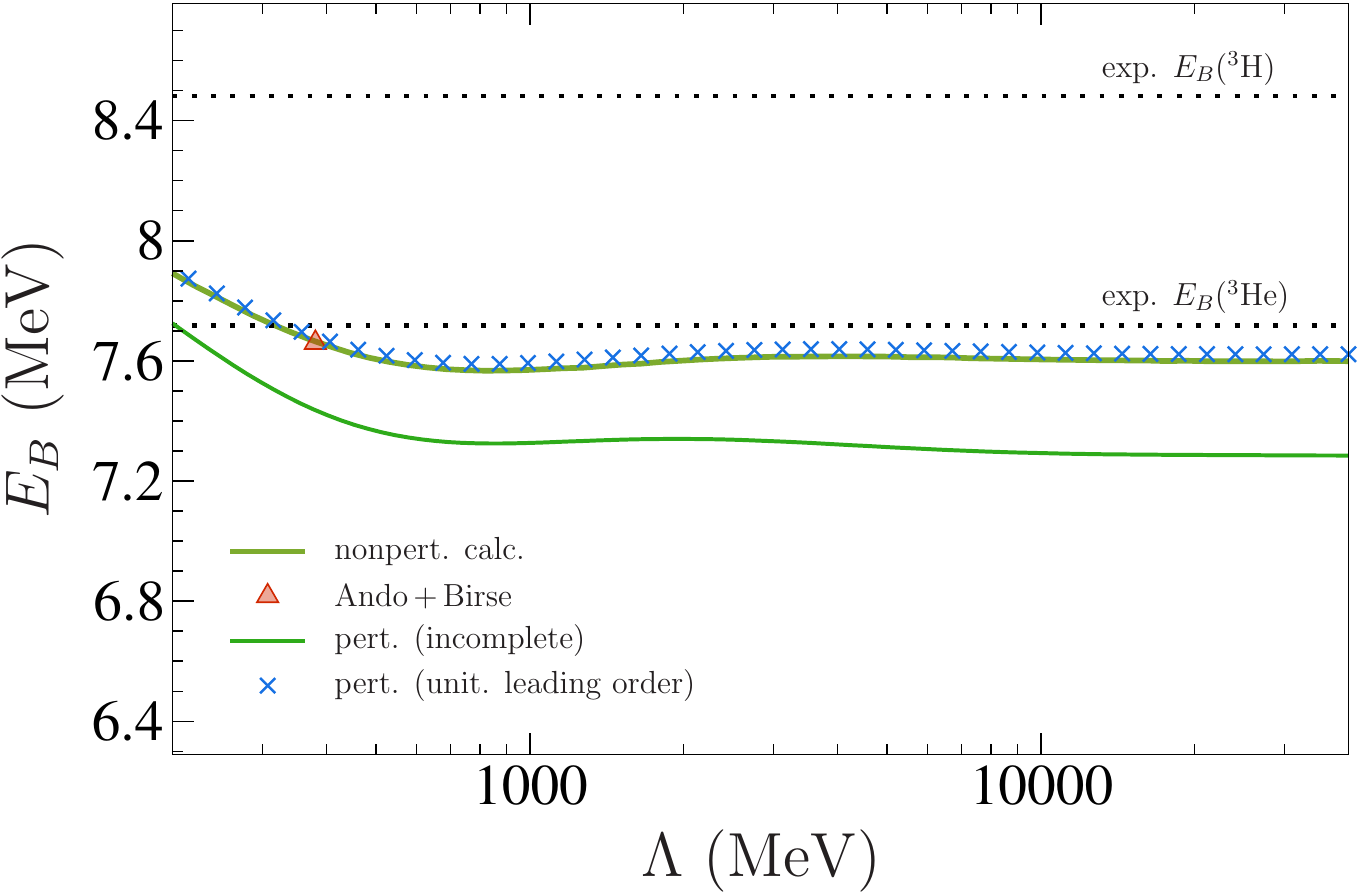}
\caption{Results for the \He binding energy (in $\MeV$) as a function of the 
cutoff (in $\MeV$).  The upper solid curve shows the result of the 
nonperturbative calculation presented in Ref.~\cite{Konig:2014ufa}; the result 
of Ref.~\cite{Ando:2010wq} is shown as a red triangle.  The incomplete 
perturbative result of Ref.~\cite{Konig:2014ufa} is given by the lower solid 
line.  Crosses represent the new complete perturbative calculation up to linear 
order in the spin-singlet scattering lengths.  For comparison, dotted horizontal 
lines indicate the experimental values for the \Triton and \He binding energies. 
The photon mass $\lambda$ has been linearly extrapolated to zero.}
\label{fig:He3-LO}
\end{figure}

To gauge the effects of perturbative Coulomb corrections, we show in 
Fig.~\ref{fig:He3-LO} the result of the new calculation presented here at NLO,  
as a function of the cutoff.  The photon mass $\lambda$ has been extrapolated to 
zero in the same way as in Ref.~\cite{Konig:2014ufa}, that is,
a linear extrapolation based on the range $\lambda = 0.4\ldots0.6~\MeV$.  All 
Coulomb effects, including those in the $pp$ sector, are included fully 
perturbatively here, meaning that we only consider matrix elements between 
trinucleon wavefunctions that involve a single Coulomb-photon exchange.  We find 
that the inclusion of the renormalized Coulomb-bubble diagram,
Fig.~\ref{fig:DeltaE-bub-C}, ensures proper renormalization of the three-body 
energy.  This is in contrast with Ref.~\cite{Vanasse:2014kxa}, which resums 
some Coulomb contributions already at LO and finds a logarithmic divergence at 
NLO when $r_C = r_t$.

The new contribution also provides a sizable (as compared to the total energy 
splitting) modification of the incomplete perturbative results of 
Ref.~\cite{Konig:2014ufa}.  It brings the full perturbative result very close to 
the non-perturbative leading-order calculation of Ref.~\cite{Konig:2014ufa}, 
which extended the results of Ref.~\cite{Ando:2010wq} to much larger cutoff 
values.  This establishes that Coulomb effects really are a completely 
perturbative correction in the \He bound state compared to the \Triton.  We 
stress that ``leading-order'' means something different in that paper than in 
the new approach presented here: Ref.~\cite{Konig:2014ufa}, as pionless 
calculations preceding it, resums certain Coulomb effects and includes the 
scattering length in the leading-order spin-singlet propagators, whereas we take 
those in the unitarity limit and only include the finite scattering lengths as 
perturbative corrections.  At the same time, our calculation thus shows that 
this is a good approximation, as expected from the $\aleph_0/Q$ expansion, which 
embodies the fact that these scattering lengths are very large.

%%%%%%%%%%%%%%%%%%%%%%%%%%%%%%%%%%%%%%%%%%%%%%%%%%%%%%%%%%%%%%%%%%%%%%%%%%%%%%
%%%%%%%%%%%%%%%%%%%%%%%%%%%%%%%%%%%%%%%%%%%%%%%%%%%%%%%%%%%%%%%%%%%%%%%%%%%%%%

At \NNLO and higher, isospin-breaking effects from the quark masses and 
higher-order electromagnetic effects will contribute.  For example,
the $\OO(\alpha^2)$ Coulomb contribution is attractive and tends to reduce
the splitting found at NLO.  Unfortunately, further shorter-range interactions
will appear.  $N\!N$ input can, to some extent, be determined from $N\!N$ data.  
Even in this case, however, there may be an isospin-breaking three-body force 
needed for proper renormalization.  At that point, one can no longer predict the 
binding energy difference, unless one can determine this force's parameter from 
another isospin-violating observable.

One of the higher-order effects comes from isospin violation in the effective 
ranges.  At our NLO there is no isospin breaking in these, so $r_t=r_C$ and 
there are no effective-range effects in the binding energy difference.  This is 
a feature by construction in our new approach: our LO state is isospin-symmetric 
in the spin-singlet channels, so perturbative corrections from isospin-symmetric 
ranges exactly cancel via the NLO adjustment of the existing three-nucleon force 
that keeps the triton in the right place.  Once one includes isospin breaking 
in the spin-singlet ranges, $r_t \neq r_C$, at some higher order, one recovers 
again the linear divergence (as a function of the ultraviolet cutoff $\Lambda$) 
that has been identified in Ref.~\cite{Vanasse:2014kxa}.

\begin{figure}[tb]
\centering
\begin{minipage}{0.49\textwidth}
 \centering
 \includegraphics[width=0.4275\textwidth,clip]{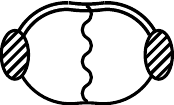}\\[0.77em]
 (a)
\end{minipage}\begin{minipage}{0.49\textwidth}
\centering
 \vspace*{0.5em}
 \includegraphics[width=0.441\textwidth,clip]{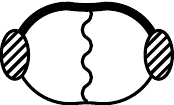}\\[0.77em]
 (b)
\end{minipage}
\caption{Combined ``Coulomb + range'' corrections contributing to the 
trinucleon binding energy in perturbation theory,
a higher-order effect in our expansion.}
\label{fig:DeltaE-sim}
\end{figure}

Further contributions that are proportional to the effective ranges come from 
the direct coupling of photons to the dibaryon fields, which are generated by 
the covariant derivatives in Eqs.~\eqref{eq:L-3S1} and~\eqref{eq:L-1S0}.  The 
corresponding diagrams are shown in Fig.~\ref{fig:DeltaE-sim}.  The expressions
for these diverge logarithmically as a function of the momentum cutoff 
$\Lambda$, as identified in 
Refs.~\cite{Koenig:2013,Vanasse:2014kxa,Konig:2014ufa}.  Essentially, the 
scaling of the diagrams in Fig.~\ref{fig:DeltaE-sim} is the same as for the 
proton bubble with a single photon exchange that we discuss in 
Sec.~\ref{sec:Coulomb}.  Whereas in that case we had momentum-independent 
vertices and a single nucleon propagator $\sim q^{-2}$ left in each loop (after 
carrying out the energy integrals), we now get a factor $1/q$ each from the 
ultraviolet behavior of the dibaryon propagators and trinucleon vertices, 
respectively.  Compared to the Coulomb correction~\eqref{eq:DeltaE-bub} the 
diagrams shown in Fig.~\ref{fig:DeltaE-sim} are suppressed by $Q/\LamNoPi$.
This relative ordering is in fact exactly the same as in previous
calculations~\cite{Vanasse:2014kxa,Konig:2014ufa,Konig:2013cia}, but in our new 
counting scheme it means that these diagrams are \NNLO.

Thus, these two divergences associated with the effective ranges appear at 
higher orders in our approach.  Note that Ref.~\cite{Vanasse:2014kxa} contains 
a third source of divergence which is linear in the effective ranges: the 
interference between non-perturbative Coulomb and perturbative range effects. 
This additional logarithmic divergence occurs because 
Ref.~\cite{Vanasse:2014kxa} employs the full Coulomb-dressed dibaryon 
propagator at leading order.  We emphasize that this divergence is absent in 
our new approach where all Coulomb contributions are treated perturbatively.  
We see here an example of the more general fact that, when singular interactions 
are involved, the cutoff dependence of perturbative diagrams is not necessarily 
the same as that of the resummed series.  As a consequence, as noted above, no 
new three-body interaction is needed for renormalization at NLO in our 
new counting.

%%%%%%%%%%%%%%%%%%%%%%%%%%%%%%%%%%%%%%%%%%%%%%%%%%%%%%%%%%%%%%%%%%%%%%%%%%%%%%
%%%%%%%%%%%%%%%%%%%%%%%%%%%%%%%%%%%%%%%%%%%%%%%%%%%%%%%%%%%%%%%%%%%%%%%%%%%%%%

The discussion here can be generalized to other isospin-violating effects.  Our 
expansion for explicit electromagnetic effects is in powers of $\alpha M_N/Q$, 
which we are counting as $\aleph_0/Q$ and pairing, for simplicity, with the 
standard pionless EFT expansion $Q/\LamNoPi$.  In addition to photon exchange, 
there are ``indirect'' electromagnetic effects that take place at short 
distances and appear in the Lagrangian as interactions among nucleons. 
Moreover, the up-down quark mass difference also generates isospin-breaking 
interactions.  The form of the interactions are dependent on the way isospin is 
broken~\cite{VanKolck:1993ee,vanKolck:1995cb}.  Since the quark masses break 
charge symmetry (a rotation of $\pi$ around the second axis in isospin space), 
their low-energy footprints will break charge symmetry as well, at least in 
first-order perturbation theory.  In contrast, electromagnetic interactions 
break isospin more generally. 

The most obvious consequence of isospin-breaking interactions is the
neutron-proton mass splitting $\delta M_N$.  We can estimate this as 
$\delta M_N=\OO(\alpha M_N/(4\pi), m_u-m_d)$,  where we included a $4\pi$ 
expected from a photon loop.  It is well known that these effects have opposite 
signs and are comparable in magnitude, but, as indicated by this estimate, the 
quark-mass contribution is somewhat larger and makes the neutron heavier.  One 
might worry that this splitting, appearing as a mass term in the Lagrangian, 
should be compared to the kinetic terms shown in Eq.~\eqref{eq:L-Nd}.  However, 
the nucleon mass splitting term can be removed by a redefinition of the nucleon 
field~\cite{Friar:2004ca}, and once this is done the splitting appears only in 
the kinetic term  itself. It is thus $\OO(\delta M_N/M_N)$ relative to leading 
order, a very small effect.

The dominant isospin-breaking effects are expected to appear in the short-range 
two-nucleon interactions, represented in Eq.~\eqref{eq:L-Nd} via dibaryon 
fields.  With the choice in Eq.~\eqref{eq:simple-y}, one counts the dibaryon
residual masses as small scales, \eg, $\sigma_t=\OO(\aleph_0)$.  To first order 
in the isospin-breaking parameters, $\sigma_{t,nn}-\sigma_t$ is proportional
to $m_u-m_d$, while $\sigma_{t,pp}-\sigma_t$ is proportional to both $(m_u-m_d)$ 
and $\alpha M_N$.  The largest sizes they are expected to have are 
$(\sigma_{t,nn}-\sigma_t)/\sigma_t = \OO((m_u-m_d)/\aleph_0)$ and 
$(\sigma_{t,pp}-\sigma_t)/\sigma_t = \OO(\alpha M_N/\aleph_0, 
(m_u-m_d)/\aleph_0)$.  The electromagnetic contribution in the $pp$ channel of
$\OO(\alpha M_N/\aleph_0) = \OO(1)$ is just the one required to renormalize 
Coulomb treated as NLO; the $\alpha \MN$ appears explicitly in
Eq.~\eqref{eq:renorm-sigma-t-pp-1}.  This counting is consistent since it yields
$a_t/a_{C}-1 = \OO(1)$, while empirical values give $a_t/a_{C}-1 \approx 2$.

How we count the quark-mass effects is a matter of choice, since $m_u-m_d$ is at 
the QCD level an independent parameter.  The estimate above suggests 
$a_t/a_{t,nn}-1 = \OO((m_u-m_d)/\aleph_0) \sim 0.3$, again consistent with the 
standard value $a_{t,nn}\simeq -18.7$~fm~\cite{GonzalezTrotter:2006wz}, which
gives $a_t/a_{t,nn}-1 \approx 0.25$.  However, there are significant 
uncertainties in the value of $a_{t,nn}$.  For example, a value $a_{t,nn} 
\simeq -16.1$~fm has also been obtained~\cite{Huhn:2001yk}, which would mean 
more significant quark mass effects.  Conversely, a value closer to $a_t$ would 
more clearly indicate $m_u-m_d$ as a separate scale, much smaller than 
$\aleph_0$.  In Ref.~\cite{Kirscher:2011zn} a pionless EFT analysis of 
the trinucleon energy splitting $\Delta E$ 
was carried out at LO in the standard power counting with an additional 
quark-mass, isospin-breaking $N\!N$ interaction.  In this case, $\Delta E$ is 
correlated with $a_{t,nn}$, and one can use the experimental value of the former 
to determine the latter.  It was found that $a_{t,nn}\simeq -(22.9\pm 4.1)$~fm, 
which gives $a_t/a_{t,nn}-1$ ranging from $-0.1$ to $0.25$.  This suggests that 
the pairing $m_u-m_d \sim \alpha M_N/4\pi$ that one could infer from the 
nucleon mass splitting can be applied to $N\!N$ interactions as well, which 
results in quark-mass effects about an order of magnitude below electromagnetic 
ones.  We therefore took in this paper the standpoint that these are \NNLO 
effects and do not contribute to the order we were working.  Once the most 
important quark mass contribution is relegated to an order where new, 
undetermined counterterms appear, it is no longer possible to constrain 
$a_{t,nn}$ from $\Delta E$~\cite{Hammer:2014rba}.

For higher terms in the $Q/\LamNoPi$ expansion, similar arguments can be used, 
but now taking into account that their parameters are determined by the
high scales as given by the pionless EFT power counting.  For 
example~\cite{vanKolck:1997ut,Kaplan:1998tg,Kaplan:1998we,vanKolck:1998bw}, 
$c_t = \OO(M_N/\LamNoPi)$, so we expect ${(c_{t,nn}-c_t)/c_t} = 
\OO((m_u-m_d)/\LamNoPi)$ and $(c_{t,pp}-c_t)/c_t = \OO(\alpha M_N/\LamNoPi, 
(m_u-m_d)/\LamNoPi)$.  These relations indicate a suppression of 
$\OO(\aleph_0/\LamNoPi)$  relative to breaking in $\sigma_{t(,\cdot\cdot)}$.
They imply $r_{C}/r_t-1 = \OO(\alpha M_N/\LamNoPi)\sim 0.05$, in agreement 
with $r_{C}/r_t-1\approx 0.02$ from empirical values.  Thus, it is consistent to 
take the electromagnetic isospin-breaking range as an \NNLO effect, with the 
prediction~\eqref{eq:rCprediction} valid up to about 5\%.  As we pointed out 
above, a linear divergence appears in the three-nucleon system which then 
requires a new, isospin-breaking three-body force at the same order.

Our result for the \He binding energy is
\begin{multline}
 \EBp(\He)^{\text{LO+NLO}} = \EBp(\Triton) + \Delta E^{\text{NLO}} \\
 = 8.48~\MeV - (0.86\pm 0.17)~\MeV
 = (7.62\pm 0.17)~\MeV \,,
\label{eq:He3-result}
\end{multline}
where we estimated the error in the energy difference as $\OO(\alpha M_N/Q, 
(m_d - m_u)/\aleph_0) \sim 20\%$---slightly larger than the ratio NLO/LO.
Eq.~\eqref{eq:He3-result} represents 98.7\% of the observed value, 
$\EBp(\He)^{\text{exp}} \approx 7.72~\MeV$.  There is room for higher-order 
contributions, but most of the splitting is accounted for at \NLO.

We emphasize that the error in Eq.~\eqref{eq:He3-result} should be understood 
as a rough estimate of higher-order contributions.  The above value comes 
from taking the mean of $a_C$ and $a_t$ for $1/\aleph_0$.  By using the 
physical value for $a_C$ as \NLO input, we are overestimating the magnitude of 
electromagnetic effects, because \NNLO corrections of both electromagnetic 
and quark-mass origin contribute to the $pp$ scattering length.  Potential-model 
calculations of photon exchange (for example, Ref.~\cite{Friar:1987zzc}) tend 
to give a $\Delta E$ consistent with an almost model-independent estimate of
about 680 keV~\cite{Brandenburg:1978aa}, but not the full isospin violation in 
the $N\!N$ scattering lengths.  Our result cannot be directly compared with 
these older perturbative-photon calculations because we include also 
shorter-range interactions---as required to ensure renormalization---that give 
the measured $pp$ scattering length.  However, the EFT allows us to directly 
study effects when $a_C$ is closer to $a_t$, effectively simulating a 
potential-model calculation with an interaction that does not give the physical
splitting in the scattering lengths.  For example we find that $a_C = -9.0\; 
(-10.0)~\fm$ leads to a trinucleon splitting of about $700$ $(600)~\keV$, and we 
can thus confirm that the older calculations are consistent with an uncertainty 
of about 20\% in $a_C$, in line with our estimate of higher-order contributions 
above.

\section{Summary and outlook}
\label{sec:Conclusion}

In this work, we have established a rearrangement of the perturbative expansion 
in pionless effective field theory that takes the spin-singlet nucleon--nucleon 
channels in the unitarity limit.  Not only does this allow us to demonstrate 
quantitatively that nature is very close to this scenario, it also enables
us to include Coulomb corrections perturbatively on the same footing.  An 
important ingredient to this is the consistent isolation of the divergent 
one-photon piece in the $pp$ sector, which then guarantees the renormalization 
of the corresponding contribution in the three-nucleon sector.

By studying the trinucleon bound-state sector, we confirm that the new 
perturbative expansion works very well: both the Phillips line and 
doublet-channel $nd$ phase shifts are barely changed by perturbative 
corrections that include the actual finiteness of these quantities, proving 
that the new leading order is a good starting point for a perturbative 
expansion.  By comparison with previous nonperturbative calculations of the 
energy splitting in the \Triton--\He iso-doublet, we furthermore show that the 
Coulomb interaction is indeed a completely perturbative effect in these bound 
states.  Our new NLO with effective-range corrections set to zero is almost on 
top of previous leading-order calculations that resum certain (but not all) 
Coulomb contributions.  The same holds when isospin-symmetric ranges are 
included.  There is no new divergence at this order, which allows us to predict
the energy splitting as $\Delta E^{\text{NLO}} = {-}(0.86\pm 0.17)$ MeV,
to be compared with the experimental value $\Delta E^{\text{exp}} \simeq 
-0.764$~MeV.

The main point of our reorganization is that, by treating small quantities in 
perturbation theory, we remove inessential parameters from the lowest orders. 
For example, a 30\% accuracy for nuclear observables susceptible to pionless 
EFT should be obtained from only two parameters at LO, with four other 
parameters at NLO ensuring a 10\% accuracy or better.  An investigation of this 
reorganization in heavier systems is our next goal.

Further improvement comes from higher orders.  Isospin-symmetric effective 
ranges do not contribute to the \Triton--\He energy difference.  The inclusion 
of isospin breaking in these effective ranges will recover the previously 
identified divergence generated by this effect; the same is true for 
contributions that are of the same order in $\alpha$ as other Coulomb effects 
included here, but suppressed further by $\rd,\rnt \sim 1/\LamNoPi$.  Confirming 
that these contributions are indeed best accounted for at \NNLO, and carrying 
out a fully consistent renormalization at this order, is beyond the scope of 
this work and relegated to future investigations.

\begin{acknowledgments}
We thank Giovanni Chirilli, Sid Coon, Dick Furnstahl, Doron Gazit, and 
Johannes Kirscher for useful discussions, and the latter also for comments on 
the manuscript.  Three of us (SK, HWG, UvK) are grateful to the organizers and 
participants of the workshop \textsc{Lattice Nuclei, Nuclear Physics and QCD -- 
Bridging the Gap} at the ECT*, Trento (Italy), for stimulating presentations and 
an atmosphere which inspired the completion of this article.  This material is 
based upon work supported in part by the NSF under Grant No. PHY--1306250 (SK), 
by the NUCLEI SciDAC Collaboration under DOE Grant DE-SC0008533 (SK), by the 
Dean's Research Chair programme of the Columbian College of Arts and Sciences
of The George Washington University (HWG), by the U.S. Department of Energy, 
Office of Science, Office of Nuclear Physics, under  Award Numbers 
DE-FG02-95ER-40907 and DE-SC0015393 (HWG) and DE-FG02-04ER41338 (UvK), by the 
BMBF under contracts 05P12PDFTE and 05P15RDFN1 (HWH), and by the Helmholtz 
Association under contract HA216/EMMI (HWH).
\end{acknowledgments}

\appendix

\section{The Coulomb bubble integrals}
\label{sec:CoulombBubbles}

\subsection{Single-photon bubble}
\label{sec:SinglePhotonBubble}

Let us consider the bubble integral with a single Coulomb-photon insertion.
The original integral is given by
\begin{equation}
 \delta I_0(k) = 4\pi\MN^2 \int\dq{q_1}\int\dq{q_2}
 \frac{1}{k^2-q_1^2+\ii\eps}
 \,\frac{1}{q_2^2+\lambda^2}
 \,\frac{1}{k^2-(\vq_1+\vq_2)^2+\ii\eps} \,.
\label{eq:SingleBubbleInt}
\end{equation}
This has been calculated by Kong and Ravndal~\cite{Kong:1999sf} using
dimensional regularization.  However, their result is not entirely correct, so 
we repeat the calculation in Sec.~\ref{sec:SinglePhotonBubble-DimReg}.  
In Sec.~\ref{sec:SinglePhotonBubble-Cutoff}, we carry out the calculation with a 
simple momentum cutoff, as used in the main part of the paper.

\subsubsection{Dimensional regularization}
\label{sec:SinglePhotonBubble-DimReg}

Following Kong and Ravndal, we set $k^2=-\gamma^2$.  Going to $d=3-\epsilon$
dimensions, and introducing the renormalization scale
$\mu$ and a Feynman parameter, we get
\begin{spliteq}
 \delta I_0 &= 4\pi\alpha\MN^2 \left(\frac\mu2\right)^{\!2\epsilon}
 \!\!\int\dqd{q_2}
 \frac{1}{q_2^2+\lambda^2}
 \int_0^1\!\dd x\!\int\dqd{q_1}
 \frac{1}{\left[x(\gamma^2+q_1^2)+(1-x)(\gamma^2+(\vq_1+\vq_2)^2)\right]^2} \\
 &= 4\pi\alpha\MN^2 \left(\frac\mu2\right)^{\!2\epsilon}
 \!\frac{\Gamma\big(\frac{1}{2}+\frac{\epsilon}{2}\big)}
 {(4\pi)^{\frac{3-\epsilon}{2}}\Gamma(2)}
 \int_0^1\!\dd x
 \int\dqd{q} \frac{1}{(q^2+\lambda^2)
 \left[(x-x^2)q^2+\gamma^2\right]^{\frac12+\frac{\epsilon}{2}}} \,.
\end{spliteq}
Further following Kong and Ravndal we set $a \equiv {\gamma^2}/{(x-x^2)}$
and introduce a second Feynman parameter to write
\begin{multline}
 \delta I_0 = 4\pi\alpha\MN^2 \left(\frac\mu2\right)^{\!2\epsilon}
 \!\frac{\Gamma\big(\frac{1}{2}+\frac{\epsilon}{2}\big)}
 {(4\pi)^{\frac{3-\epsilon}{2}}\,\Gamma(2)}
 \int_0^1\!\dd x \frac{1}{(x-x^2)^{\frac12+\frac{\epsilon}{2}}} \\
 \times\int_0^1\!\dd\omega\int\dqd{q}
 \frac{\omega^{-\frac12+\frac\epsilon2}}{\left[(1-\omega)(q^2+\lambda^2)
 + \omega(q^2 + a)\right]^{\frac32+\frac{\epsilon}{2}}} \,.
\end{multline}
Noting that
\begin{equation}
 (1-\omega)(q^2+\lambda^2) + \omega(q^2 + a)
 = q^2 + (1-\omega)\lambda^2 + \omega \frac{\gamma^2}{x-x^2} \,,
\end{equation}
we can carry out the integral and arrive at
\begin{equation}
 \delta I_0 = 4\pi\alpha\MN^2 \left(\frac\mu2\right)^{\!2\epsilon}
 \!\!\!\frac{\Gamma\big(\frac{1}{2}+\frac{\epsilon}{2}\big)\Gamma(\epsilon)}
 {(4\pi)^{3-\epsilon}\,\Gamma(2)\Gamma\big(\frac32+\frac\epsilon2\big)}
 \int_0^1\!\!\dd x \frac{1}{(x-x^2)^{\frac12+\frac{\epsilon}{2}}}
 \int_0^1\!\!\dd\omega\,\frac{\omega^{-\frac12+\frac\epsilon2}}
 {\left((1-\omega)\lambda^2+\omega\dfrac{\gamma^2}{x-x^2}\right)^{\!\!\epsilon}}
 \,.
\end{equation}
Let us consider first the case $\lambda=0$, as done by Kong and Ravndal.  In
that case, the two integrals are
\begin{multline}
 \int_0^1\!\dd x \frac{1}{(x-x^2)^{\frac12+\frac{\epsilon}{2}}}
 \int_0^1\!\dd\omega\,\frac{\omega^{-\frac12+\frac\epsilon2}}
 {\left(\omega\dfrac{\gamma^2}{x-x^2}\right)^{\!\!\epsilon}}
 = \gamma^{-2\epsilon}
 \int_0^1\!\dd x\,(x-x^2)^{-\frac12+\frac{\epsilon}{2}}
 \int_0^1\!\dd\omega\,\omega^{-\frac12-\frac\epsilon2} \\
 = \gamma^{-2\epsilon}\,\frac{2^{-\epsilon}\sqrt{\pi}
 \,\Gamma\big(\frac{1}{2}+\frac{\epsilon}{2}\big)}{\Gamma\big(1+\frac{\epsilon}{2}\big)}
 \frac{2}{1-\epsilon} \,.
\end{multline}
Hence,
\begin{multline}
 \delta I_0^{\lambda=0} = 
 \frac{\alpha\MN^2}{8\pi^{3/2}} 
 \left(\frac{\pi \mu^2}{2\gamma^2}\right)^{\!\epsilon}
 \frac{\left[\Gamma\big(\frac{1}{2}+\frac{\epsilon}{2}\big)\right]^2
  \Gamma(\epsilon)}
 {\Gamma(2)\,\Gamma\big(\frac32+\frac\epsilon2\big)
  \,\Gamma\big(1+\frac{\epsilon}{2}\big)}
 \,\frac{1}{1-\epsilon} \\
 = \frac{\alpha\MN^2}{4\pi}\left[\frac{1}{\epsilon}
 + 2\log\frac{\mu\sqrt{\pi}}{2\gamma} - \EulerGamma\right] + \OO(\epsilon) \,.
\label{eq:I0-DR-final}
\end{multline}
Note that this differs slightly from the result obtained by Kong and Ravndal
due to an error\footnote{The argument of the Gamma function in the denominator 
of Eq.~(100) in Ref.~\cite{Kong:1999sf} should be $3-d/2$ instead of $2-d/2$.} 
in Ref.~\cite{Kong:1999sf}.  If we keep the photon-mass term, the integrals 
over the Feynman parameters cannot be separated.  After a couple of steps one 
arrives at
\begin{multline}
 \int_0^1\!\dd x \frac{1}{(x-x^2)^{\frac12+\frac{\epsilon}{2}}}
 \int_0^1\!\dd\omega\,\frac{\omega^{-\frac12+\frac\epsilon2}}
 {\left((1-\omega)\lambda^2+\omega\dfrac{\gamma^2}{x-x^2}
 \right)^{\!\!\epsilon}} \\
 = 2\pi + 2\pi\,\epsilon
 \left[1+ 2\log\frac{\mu}{2\gamma}
 + 2\log\!\left(1+\frac{\lambda}{2\gamma}\right)\right]
 + \OO(\epsilon^2) \,,
\end{multline}
so eventually we get
\begin{equation}
 \delta I_0 = \frac{\alpha\MN^2}{4\pi}\left[\frac{1}{\epsilon}
 + 2\log\frac{\mu\sqrt{\pi}}{2\gamma}
 - 2\log\!\left(1+\frac{\lambda}{2\gamma}\right)
 - \EulerGamma\right] + \OO(\epsilon) \,,
\end{equation}
which reduces to Eq. \eqref{eq:I0-DR-final} when $\lambda=0$.

\subsubsection{Simple momentum cutoff}
\label{sec:SinglePhotonBubble-Cutoff}

We now consider the integral with a simple momentum cutoff $\Lambda$.  In that
case, Eq.~\eqref{eq:SingleBubbleInt} (with $\lambda=0$) becomes
\begin{equation}
 \delta I_0^{\lambda=0}(k)
 = 4\pi\alpha\MN^2 \int^\Lambda\dq{q_1}\int^\Lambda\dq{q_2}
 \frac{1}{k^2-q_1^2+\ii\eps}
 \,\frac{1}{q_2^2}
 \,\frac{1}{k^2-(\vq_1+\vq_2)^2+\ii\eps} \,,
\label{eq:SingleBubbleInt-cutoff}
\end{equation}
or, more conveniently for the present calculation,\footnote{The shift involved 
in this rearrangement, which is equivalent to a different momentum assignment 
in the corresponding Feynman diagram, should only affect the terms suppressed 
by inverse powers of the cutoff.}
\begin{equation}
 \delta I_0^{\lambda=0}(k)
 = 4\pi\alpha\MN^2 \int^\Lambda\dq{q_1}\int^\Lambda\dq{q_2}
 \frac{1}{k^2-q_1^2+\ii\eps}
 \,\frac{1}{(\vq_1-\vq_2)^2}
 \,\frac{1}{k^2-q_2^2+\ii\eps} \,.
\label{eq:SingleBubbleInt-cutoff-2}
\end{equation}
This is logarithmically divergent, so based on dimensional analysis we know that
\begin{equation}
 \delta I_0^{\lambda=0}(k) = \frac{\alpha\MN^2}{4\pi}
 \left[A\,\log\frac{\Lambda}{k} + B + \OO(k/\Lambda)\right] \,,
\end{equation}
with some dimensionless constants $A$ and $B$.  To determine these, we first 
carry out the angular integrals in Eq.~\eqref{eq:SingleBubbleInt-cutoff-2} (all 
except one of which are trivial).  As done in the calculation with dimensional 
regularization, we also analytically continue to $k = \ii\gamma$, 
$\gamma>0$ and let $\eps\to0$.  This gives
\begin{equation}
 \delta I_0^{\lambda=0}(k) = \frac{\alpha\MN^2}{\pi^2}
 \int_0^\Lambda\dd q_1\,q_1^2 \int_0^\Lambda\dd q_2\,q_2^2
 \,\frac{\log\left[(q_1^2+q_2^2+2q_1q_2)/(q_1^2+q_2^2-2q_1q_2)\right]}
 {2q_1q_2 (q_1^2+\gamma^2) (q_2^2+\gamma^2)} \,.
\end{equation}
Carrying out the integration over $q_1$ gives
\begin{multline}
 \delta I_0^{\lambda=0}(k) = {-}\frac{\alpha\MN^2}{2\pi^2}
 \int_0^\Lambda \frac{\dd q\,q}{q^2+\gamma^2}
 \times \Bigg[\log\!\left(\frac{\Lambda-q}{\Lambda+q}\right)
 \log\!\left(\frac{\Lambda^2+\gamma^2}{q^2+\gamma^2}\right) \\
 \null + \LiTwo\!\left(\frac{q-\Lambda}{q-\ii\gamma}\right)
 - \LiTwo\!\left(\frac{q+\Lambda}{q-\ii\gamma}\right)
 + \LiTwo\!\left(\frac{q-\Lambda}{q+\ii\gamma}\right)
 - \LiTwo\!\left(\frac{q+\Lambda}{q+\ii\gamma}\right)
 \Bigg] \,,
\end{multline}
where we have renamed the second loop momentum $q_2 \rightarrow q$.  The 
remaining integral can be carried out using an indefinite integral which is 
given by a complicated sum of logarithms and polylogarithms.  We do not write 
this out here, but simply note that the final result can be asymptotically 
expanded for large $\Lambda$ to yield
\begin{equation}
 A = 1
 \mathtext{and}
 B = {-}\left(\log2 + \frac{7\,\zeta(3)}{2\pi^2}\right) \,,
\end{equation}
so that overall
\begin{equation}
 \delta I_0^{\lambda=0} = \frac{\alpha\MN^2}{4\pi}
 \left[\log\frac{\Lambda}{\gamma} - C_\zeta + \OO(k/\Lambda)\right] \,,
\label{eq:delta-I0prime}
\end{equation}
where for convenience we have defined
\begin{equation}
 C_\zeta = \log2 + \frac{7\,\zeta(3)}{2\pi^2} \approx 1.11943.
\end{equation}
Adding $0 = \log(\alpha\MN/2) - \log(\alpha\MN/2)$ to 
Eq.~\eqref{eq:delta-I0prime} we arrive at Eq.~\eqref{eq:delta-I0}.

\subsection{Fully dressed bubble}
\label{sec:FullyDressedBubble-KR}

The fully dressed $pp$ bubble shown in Fig.~\ref{fig:DressedBubble} resums 
the contribution of an arbitrary number of Coulomb-photon exchanges.  We 
discuss it here in a number of schemes, starting with the results of Kong and 
Ravndal~\cite{Kong:1999sf}, who used both dimensional regularization and a 
simple momentum cutoff.  As discussed in the main text, Ref.~\cite{Kong:1999sf}
splits up the full bubble integral into a finite piece that gives the function 
$H(\eta)$ in the Coulomb-modified effective range expansion, and a divergent 
piece, which is the only integral that is regularized.  In dimensional 
regularization, Kong and Ravndal find, for $d=3-\epsilon$,
\begin{multline}
 J_0^{\text{div}} =
 {-}\MN \left(\frac{\mu}{2}\right)^{\!\epsilon}
 \frac{2\pi^{d/2}}{(2\pi)^d\,\Gamma(d/2)}
 \int_0^\infty \dd q \, q^{d-3}
 \,\frac{2\pi\eta(q)}{\ee^{2\pi\eta(q)}-1}\frac{1}{q^2} \\
 = \frac{\alpha\MN^2}{4\pi}
 \left(\frac1\epsilon + \log\frac{\mu\sqrt{\pi}}{\alpha\MN} + 1
 - \frac32\EulerGamma\right) - \frac{\mu\MN}{4\pi} \,,
\label{eq:J0-div-DR-final}
\end{multline}
where the result in the second line is obtained through the substitution $x = 
2\pi\eta(q)$, which leads to the integral
\begin{equation}
 \int_0^1 \dd x\,\frac{x^{\epsilon-1}}{\ee^x-1}
 = \Gamma(\epsilon)\zeta(\epsilon) \,.
\label{eq:Zeta-int}
\end{equation}
A subsequent expansion of the Zeta and Gamma functions gives the PDS pole 
in $d=2$ and the $1/\epsilon$ divergence for $d\to3$, respectively.  With a 
simple cutoff, the result is given in Eq.~\eqref{eq:J0-div-Cutoff}.

Note that Kong and Ravndal only consider the full series (empty bubble, one 
photon exchange, two photon exchanges, \textit{etc.}) shown in 
Fig.~\ref{fig:DressedBubble} at once, although the logarithmic divergence really 
comes only from the single-photon contribution.\footnote{In addition, the PDS 
pole in $d=2$ corresponds to the linear divergence of the empty bubble.}
This leads to the puzzling result that while the $1/\epsilon$ divergences for 
the single-photon and fully dressed bubbles (see Eqs.~\eqref{eq:I0-DR-final} 
and~\eqref{eq:J0-div-DR-final}, respectively) agree (same prefactor/residue), 
the logarithmic pieces that depend on the renormalization scale $\mu$, as well 
as the constant terms, do not.  Such an effect can be created by splitting up a 
formally divergent expression and applying the regularization only to a part of 
it.  To see this issue, note for example that the integrand in 
Eq.~\eqref{eq:J0-div-DR-final} cannot be expanded to order $\alpha$ to recover 
the single-photon result.

While in Ref.~\cite{Kong:1999sf} this does not actually play a role---although 
they discuss the single-photon bubble for illustration, the authors eventually 
only use the fully dressed bubble for the matching to the Coulomb-modified 
scattering length---in the current work we require a scheme that is consistent 
throughout, \ie, we want to isolate the divergent single-photon piece within the 
fully dressed bubble.  In Sec.~\ref{sec:CoulomBubbles-ConfigSpace} we first 
establish in configuration space that the logarithmically-divergent pieces 
should indeed be the same for both bubble diagrams, before we finally discuss 
in Sec.~\ref{sec:FullyDressedBubble-T} our approach of using the Coulomb 
T-matrix to isolate the the single-photon contribution, while still giving a 
closed expression for the rest.

\subsection{Configuration space}
\label{sec:CoulomBubbles-ConfigSpace}

Let us consider the two bubble integrals in configuration space in order to 
analyze their divergences more carefully.  To this end, note that what we 
really want to calculate is the zero-to-zero Green's
functions~\cite{Kong:1999sf,Kong:2000px},
\begin{equation}
 J_0(k) = G_C(E;\vecr'=\vZero,\vecr=\vZero) \mathtext{,} E = k^2/\MN \,.
\end{equation}
Of course, this is infinite, so we want to isolate the divergence by starting
from a finite separation and considering the case where it goes to zero.
This procedure only deals with well-defined finite quantities, so we expect it
to avoid subtleties associated with splitting up formally infinite terms.

The Coulomb Green's function satisfies the equation
\begin{equation}
 \frac{1}{2\mu}\left(\Laplace_r + \frac{2\mu\alpha}{r} + k^2\right)
 G_C(E;\vecr,\vecr') = \vdelta(\vecr-\vecr') \,,
\label{eq:GC-GF}
\end{equation}
in exact correspondence to the free case,
\begin{equation}
 \frac{1}{2\mu}\big(\Laplace_r + k^2\big)\,G_0^{(+)}(E;\vecr,\vecr')
 = \vdelta(\vecr-\vecr') \,.
\label{eq:G0-GF}
\end{equation}
For the present case, the reduced mass is $\mu=\MN/2$.  
This Green's function is known
analytically~\cite{Hostler:1963zz,Meixner:1933aa}
when one argument is set to zero.  Adjusting to agree with the
conventions used in Ref.~\cite{Koenig:2013}, we have
\begin{equation}
 G_C(E;\vecr'=\vZero,\vecr)
 = {-}2\mu\frac{\Gamma(1-\ii\eta)\,W_{\ii\eta;1/2}(-2\ii k r)}{4\pi r}
 \mathtext{,} \eta = \frac{\alpha\mu}{k} \,.
\label{eq:GC-W}
\end{equation}
Expanding first the Whittaker function around $r=0$, and subsequently the
whole right-hand side in powers of $\alpha$, we find
\begin{equation}
 G_C(E;\vecr'=\vZero,\vecr) = {-}\frac{\MN}{4\pi}\left[\frac{1}{r}
 + \ii k + \alpha\MN\Big(1 - \EulerGamma - \log(-2\ii kr)\Big)\right]
 + \OO(r) \,.
\label{eq:GC-0}
\end{equation}
The $1/r$ and $\log(r)$ terms in the above expression directly correspond to
the linear and logarithmic divergences that we found previously in momentum
space.

The free Green's function, \ie, the solution to Eq.~\eqref{eq:G0-GF}, is given 
by
\begin{equation}
 G_0^{(+)}(E;\vecr,\vecr') = \mbraket{\vecr}{\hat{G}_0^{(+)}(E)}{\vecr'}
 = {-\frac{\mu}{2\pi}}\frac{\ee^{\ii k|\vecr-\vecr'|}}{|\vecr-\vecr'|} \,.
\label{eq:G0-r}
\end{equation}
The first two terms in Eq. \eqref{eq:GC-0} come from, with $\mu=\MN/2$,
\begin{equation}
 G_0^{(+)}(E;\vecr'=0,\vecr) = 
 {-\frac{\MN}{4\pi}}\left[\frac{1}{r} + \ii k \right] + \OO(r) \,.
\label{eq:G0-r-expanded}
\end{equation}
This corresponds to the contribution of the empty bubble.

We now turn to the single-photon diagram and expect to find exactly the 
remaining divergence, because diagrams with two and more photons should be 
finite.  What we have to calculate is thus
\begin{equation}
 G_C^{(1)}(E;\vecr',\vecr)
 = \mbraket{\vecr}{\hat{G}_0^{(+)}(E)\,\hat{V_C}\,\hat{G}_0^{(+)}(E)}{\vecr'}
 \,,
\label{eq:GC1-1}
\end{equation}
where 
\begin{equation}
 \mbraket{\vecr}{\hat{V_C}}{\vecr'}
 = \vdelta(\vecr-\vecr')\,\frac{\alpha}{r}
\end{equation}
is the Coulomb potential.  Setting $k = \sqrt{\mathstrut2\mu E+\ii\eps}$ and 
$\vecr'=\vZero$ in Eq.~\eqref{eq:GC1-1} gives
\begin{equation}
 G_C^{(1)}(E;\vZero,\vecr) = \frac{\alpha\mu^2}{2\pi}
 \int_0^\infty\dd r''\,\ee^{\ii k r''}
 \int_{-1}^1\dd\!\cos\theta\,\frac{\ee^{\ii k\abs{\vecr-\vecr''}}}
 {\abs{\vecr-\vecr''}} \mathtext{,} \vecr''\cdot\vecr = r''\,r\cos\theta \,,
\label{eq:GC1-2}
\end{equation}
where we have inserted identities to separate the operators and carried out the
trivial integrals.  The angular integral simply recovers the well-known
expression for the $S$-wave projected free Green's function:
\begin{equation}
 \int_{-1}^1\dd\!\cos\theta\,\frac{\ee^{\ii k\abs{\vecr-\vecr''}}}
 {\abs{\vecr-\vecr''}}
 = {-}\frac{2}{k}\,\frac{\ee^{\ii k r_>} \sin(kr_<)}{r''\,r}
 \mathtext{,} r_> = \max(r,r'') \mathtext{,} r_< = \min(r,r'') \,.
\end{equation}
The remaining integral in Eq.~\eqref{eq:GC1-2} can then be done by splitting up
the domain according to
\begin{equation}
 \int_0^\infty\dd r'' = \int_0^r\dd r'' + \int_r^\infty\dd r'' \,.
\end{equation}
In each region, $r_<$ and $r_>$ are then uniquely defined.  The first integral
is
\begin{equation}
 \int_0^r \frac{\ee^{\ii k r''}\sin(kr'')}{r''} \dd r''
 = \frac{\ii}{2}\Big(\EulerGamma + \log(2kr) - \Ci(2kr)
 - \ii \; \Si(2kr)\Big) \,,
\end{equation}
in terms of the sine and cosine integrals~\cite{Gradshteyn:2007-SiCi}, 
respectively
\begin{subalign}
 \Si(z) &= \int_0^z \frac{\sin(t)}{t} \dd t = \OO(z) \,, \\
 \Ci(z) &= {-}\!\int_z^\infty \frac{\cos(t)}{t} \dd t
 = \EulerGamma + \log z + \OO(z^2) \,.
\end{subalign}
For the second integral we find
\begin{equation}
 \int_r^\infty \frac{\ee^{2\ii k r''}}{r''} \dd r'' = \Gamma(-2\ii k r) \,.
\end{equation}
Combining the results, we get
\begin{equation}
 G_C^{(1)}(E;\vecr'=\vZero,\vecr) = {-}\frac{\alpha\mu^2}{\pi kr}\left\{\!
 \frac{\ii\ee^{\ii k r}}{2}
 \Big[\EulerGamma+\log(2kr)-\Ci(2kr)-\ii\,\Si(2kr)\Big]
 + \sin(kr) \Gamma(-2\ii k r) \!\right\} \,.
\label{eq:GC1-3}
\end{equation}
Finally, expanding this around $r=0$ and inserting $\mu=\MN/2$, we arrive at
\begin{equation}
 G_C^{(1)}(E;\vecr'=\vZero,\vecr) = {-}\frac{\alpha\MN^2}{4\pi}\Big[
 1 - \EulerGamma - \log(-2\ii kr)\Big] + \OO(r) \,.
\label{eq:GC1-0}
\end{equation}
This is exactly the term linear in $\alpha$ in Eq.~\eqref{eq:GC-0}.

\subsection{Fully dressed bubble with T-matrix expansion}
\label{sec:FullyDressedBubble-T}

Having established that the remaining divergence indeed comes exactly from the
single-photon diagram, let us take another shot at consistently isolating it in
a momentum-space calculation.  To this end, note that the Coulomb Green's
function satisfies the operator equation
\begin{equation}
 \hat{G}_C(E) = \hat{G}_0^{(+)}(E)
 + \hat{G}_0^{(+)}(E)\,\hat{V}_C\,\hat{G}_C(E) \,.
\label{eq:GC-LS}
\end{equation}
On the other hand, we have the Coulomb T-matrix defined by the
Lippmann--Schwinger equation
\begin{equation}
 \hat{T}_C(E) = \hat{V}_C + \hat{V}_C\,\hat{G}_0^{(+)}(E)\,\hat{T}_C(E) \,.
\label{eq:LS-C}
\end{equation}
With
\begin{equation}
 \hat{V}_C\,\hat{G}_C(E) = \hat{T}_C(E)\,\hat{G}_0^{(+)}(E) \,,
\end{equation}
we can write Eq.~\eqref{eq:GC-LS} as
\begin{equation}
 \hat{G}_C(E) = \hat{G}_0^{(+)}(E)
 + \hat{G}_0^{(+)}(E)\,\hat{T}_C\,\hat{G}_0^{(+)}(E) \,.
\label{eq:GC-LS-2}
\end{equation}
The advantage of this operation becomes apparent once we note that the Coulomb
T-matrix is known in closed form.  For example, in terms of the Coulomb 
potential in momentum space
\begin{equation}
 V_C(\vecp,\vecq) = \frac{4\pi\alpha}{(\vp-\vq)^2}\,,
\end{equation}
one has
\begin{equation}
 T_C(k;\vecp,\vecq) = \mbraket{\vecp}{\hat{T}_C}{\vecq}
 = V_C(\vecp,\vecq) \left\{1-2\ii\eta
 \int_1^\infty\left(\frac{s+1}{s-1}\right)^{\!\!-\ii\eta}
 \frac{\dd s}{s^2-1-\epsilon} \right\} \,,
\label{eq:TC-int}
\end{equation}
or
\begin{multline}
 T_C(k;\vecp,\vecq) = V_C(\vecp,\vecq) \Bigg\{1-\Delta^{-1}\Big[
 {_2F_1}\!\left(1,\ii\eta,1+\ii\eta;\frac{\Delta-1}{\Delta+1}\right) \\
 -{_2F_1}\!\left(1,\ii\eta,1+\ii\eta;\frac{\Delta+1}{\Delta-1}\right)
 \Big]\Bigg\} \,,
\label{eq:TC-hyp}
\end{multline}
with
\begin{equation}
 \epsilon = \frac{(p^2-k^2)(q^2-k^2)}{k^2(\vecp-\vecq)^2}
 \mathtext{,} \Delta^2 = 1 + \epsilon \,.
\end{equation}
For these and a number of alternative representations, see for example
Ref.~\cite{Chen:1972ab}.  Note that both Eq.~\eqref{eq:TC-int}
and~\eqref{eq:TC-hyp} express the T-matrix as the Coulomb potential plus a
remainder that summarizes the exchange of two and more photons,
\begin{equation}
 T_C(k;\vecp,\vecq) = V_C(\vecp,\vecq) + T_C^\Delta(k;\vq,\vp)\,.
\end{equation}
Hence, we can write the fully resummed bubble integral as
\begin{equation}
 J_0(k) = I_0(k) + \delta I_0(k) + \delta J_0(k) \,,
\label{eq:J0-split}
\end{equation}
where
\begin{equation}
 I_0(k) = \MN\int\ddq \frac{1}{k^2-q^2+\ii\eps}
\end{equation}
is the empty bubble, $\delta I_0(k)$ is exactly the single-photon bubble
calculated in Sec.~\ref{sec:SinglePhotonBubble}, and $\delta J_0(k)$
remains to be investigated.  The crucial point is that it is supposed to be 
ultraviolet finite.

The integral we have to evaluate is
\begin{equation}
 \delta J_0(k) = \MN^2\int\dq{p}\int\dq{q} 
 \frac{T_C^\Delta(k;\vq,\vp)}{(p^2-k^2-\ii\eps)(q^2-k^2-\ii\eps)} \,.
\end{equation}
Noting that $T_C^\Delta(k;\vq,\vp)$ depends only on the angle between
$\vq$ and $\vp$, we can carry out all but one angular integral.  The remaining
one over $\cos\theta$ is then an $S$-wave projection.  Fortunately, partial-wave
projections of the full Coulomb T-matrix are known in closed
form~\cite{Dusek:1981aa,vanHaeringen:1975cz}. The $S$-wave result can be written
as~\cite{vanHaeringen:1975cz}
\begin{multline}
 T_{C,0}(E=k^2/(2\mu);p,q) = \frac{\ii\pi}{\mu} \frac{k}{pq}
 \Big[
 {_2F_1}\!\left(1,\ii\eta,1+\ii\eta;a'a\right)
 - {_2F_1}\!\left(1,\ii\eta,1+\ii\eta;a/a'\right) \\
 + {_2F_1}\!\left(1,\ii\eta,1+\ii\eta;1/(a'a)\right)
 - {_2F_1}\!\left(1,\ii\eta,1+\ii\eta;a'/a\right)
 \Big] \,,
\end{multline}
where
\begin{equation}
 a = \frac{p-k}{p+k} \mathtext{,} a' = \frac{q-k}{q+k} \,.
\end{equation}
With the projection of the Coulomb potential alone,
\begin{equation}
 V_{C,0}(p,q) = \frac{2\pi\alpha}{pq} Q_0\!\left(\frac{p^2+q^2}{2pq}\right)
 = \frac{\pi\alpha}{pq}\log\!\left(\frac{2pq+p^2+q^2}{2pq-p^2-q^2}\right) \,,
\end{equation}
we can write
\begin{equation}
 \delta J_0(k) = \left(\frac\MN{2\pi^2}\right)^{\!2}
 \int\dd p\,p^2 \int\dd q\,q^2
 \,\frac{T_{C,0}(E=k^2/(2\mu);p,q) - V_{C,0}(p,q)}
 {(p^2-k^2-\ii\eps)(q^2-k^2-\ii\eps)} \,.
\label{eq:deltaJ0-int}
\end{equation}

\begin{figure*}[tb]
 \centering
 \begin{overpic}[width=0.495\textwidth]{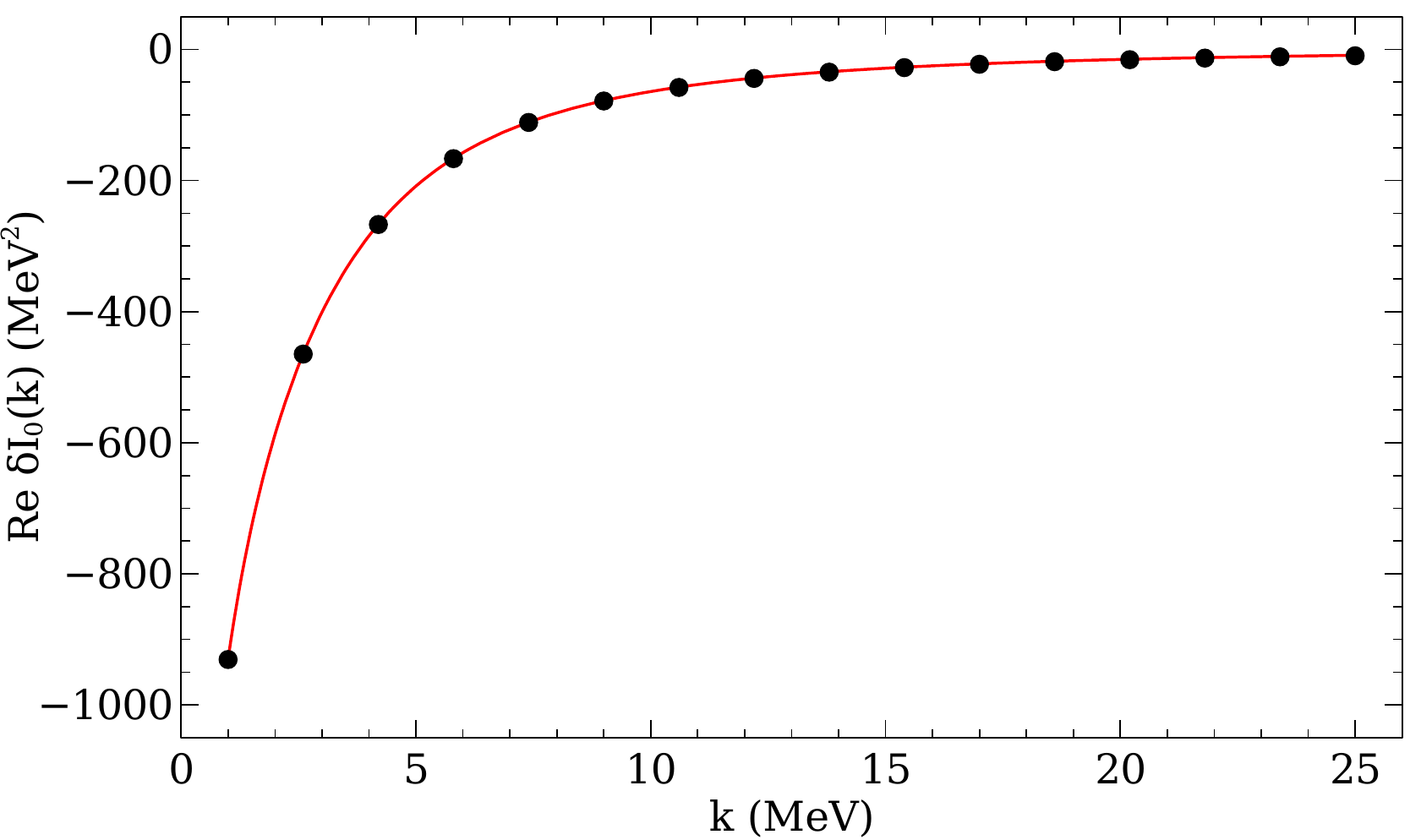}
  \put(87.5,12){\textcolor{black}{\large (a)}}
 \end{overpic}~~~\begin{overpic}[width=0.495\textwidth]{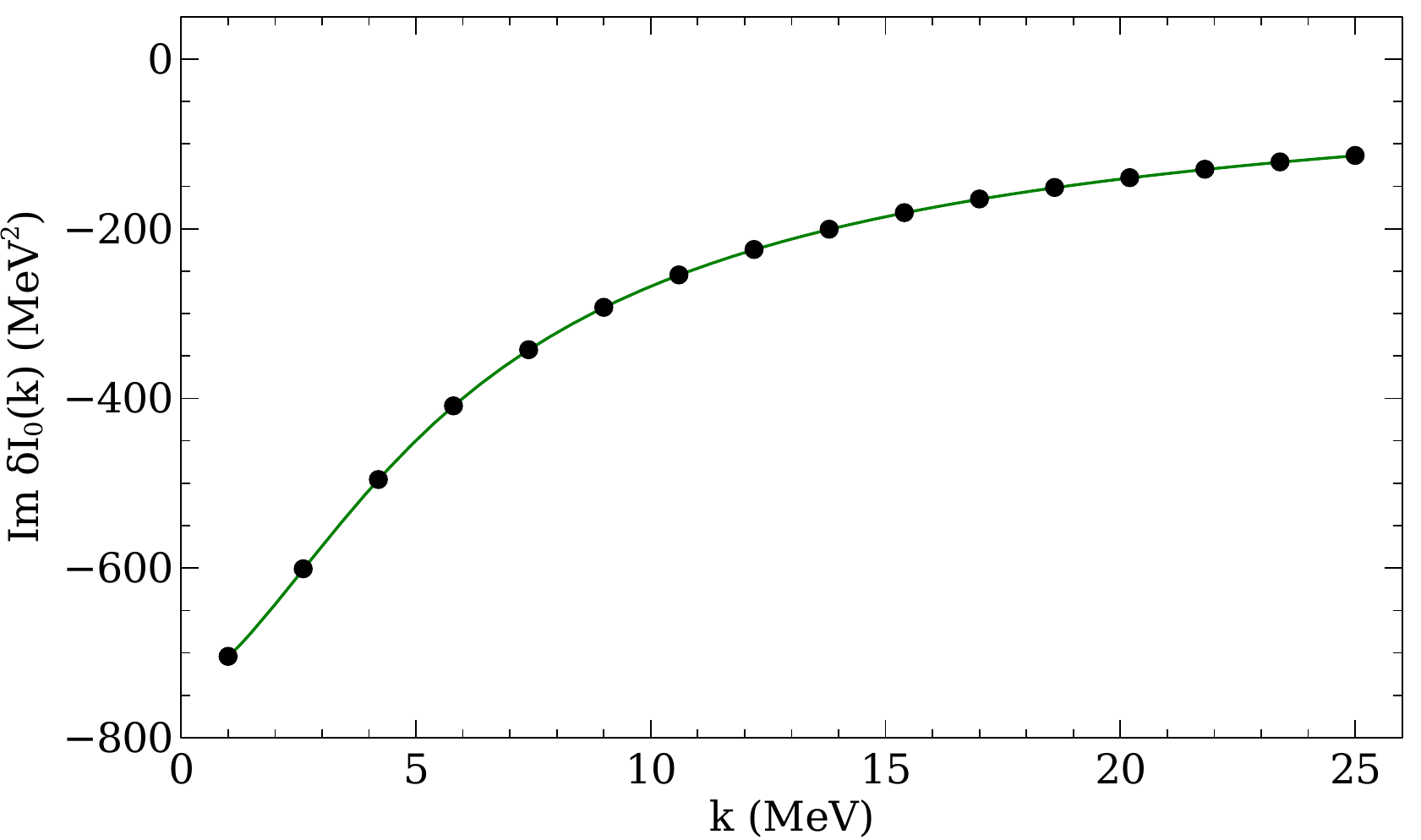}
  \put(87.5,12){\textcolor{black}{\large (b)}}
 \end{overpic}
 \caption{Numerical evaluation of $\delta J_0(k)$:
(a) real part; (b) imaginary part.  Black dots correspond to the integral given 
in Eq.~\eqref{eq:deltaJ0-int}.  The solid lines show the functions defined in 
Eqs.~\eqref{eq:deltaJ0-num-re} (a) and~\eqref{eq:deltaJ0-num-im} (b).
The calculation was performed with an explicit small imaginary part of 
$\eps=0.05~\MeV$ added to the real momentum $k$ and a momentum cutoff of 
$16000~\MeV$.}
\label{fig:J0-compare}
\end{figure*}

This integral looks difficult to solve, but at the very least it can be
evaluated numerically.  We do this with a (large) momentum cutoff and a small 
imaginary part of $k$ in place for regularization.  Considering the real part, 
one finds that it is consistent with
\begin{equation}
 \Rp\,\delta J_0(k) = {-}\frac{\alpha\MN^2}{4\pi}
 \left(\Rp\left\{\psi(\ii\eta) + \frac{1}{2\ii\eta}\right\} + C_\Delta\right)
\label{eq:deltaJ0-num-re}
\end{equation}
with a constant
\begin{equation}
 C_\Delta \approx 0.579 \,.
\label{eq:C-Delta}
\end{equation}
The level of agreement is demonstrated in Fig.~\ref{fig:J0-compare}(a).  Note 
that Eq.~\eqref{eq:deltaJ0-num-re} contains the function $H(\eta)$ except 
for the $\log\ii\eta$ part.  This agrees with what one would expect here because
in the sum~\eqref{eq:J0-split} we already have a $\log k$ term from the 
single-photon contribution $\delta I_0(k)$.  Considering the imaginary part, we 
find that it agrees very well with
\begin{equation}
 \Ip\,\delta J_0(k) = {-}\frac{\alpha\MN^2}{4\pi}
 \Ip\left\{\psi(\ii\eta) + \frac{1}{2\ii\eta}\right\}
 + \frac{\MN k}{4\pi} \,,
\label{eq:deltaJ0-num-im}
\end{equation}
as shown in Fig.~\ref{fig:J0-compare}(b).  Altogether, we have thus established 
numerically that
\begin{equation}
 \delta J_0(k) = {-}\frac{\alpha\MN^2}{4\pi}
 \left(\psi(\ii\eta) + \frac{1}{2\ii\eta} + C_\Delta\right)
 + \frac{\MN}{4\pi} \ii k \,.
\end{equation}
We expect that increased numerical accuracy would yield 
$C_\Delta\to\EulerGamma\approx0.577216$.  The value in Eq.~\eqref{eq:C-Delta} 
represents our most accurate result (obtained with a momentum cutoff of 
$16000~\MeV$ and $\Ip k=0.05~\MeV$); decreasing the former and increasing the 
latter gives larger values for $C_\Delta$, so indeed we see a convergence 
pattern.

%%%%%%%%%%%%%%%%%%%%%%%%%%%%%%%%%%%%%%%%%%%%%%%%%%%%%%%%%%%%%%%%%%%%%%%%%%%%%%%%
%%%%%%%%%%%%%%%%%%%%%%%%%%%%%%%%%%%%%%%%%%%%%%%%%%%%%%%%%%%%%%%%%%%%%%%%%%%%%%%%

\end{document}